\documentclass{article}
\usepackage{chicago}
\usepackage{amssymb}
\usepackage{tabularx}

\newcommand{\ie}{\mbox{i.\,e.\,\ }}
\newcommand{\iec}{\mbox{i.\,e.\,}}

\newcommand{\egc}{\mbox{e.\,g.\,}}

\newcommand{\vctr}[1]{\ensuremath{\mathbf{ #1 }}}

\newcommand{\dr}[1]{\ensuremath{\mathrm{d} #1\,}}
\newcommand{\mc}[1]{\ensuremath{\mathcal{#1}}}

\newcommand{\pbp}[2]{\ensuremath{\frac{\partial #1}{\partial #2}}}

\newcommand{\vbv}[2]{\ensuremath{\frac{\delta #1}{\delta #2}}}

\newcommand{\ket}[1]{\ensuremath{\left|  #1 \right\rangle}}
\newcommand{\bra}[1]{\ensuremath{\left\langle #1 \right|}}
\newcommand{\bk}[2]{\ensuremath{\left\langle #1 | #2 \right\rangle}}

\newcommand{\matel}[3]{\ensuremath{\bra{#1} #2 \ket{#3}}}
 
\newcommand{\op}[1]{\ensuremath{\widehat{\textsf{\ensuremath{#1}}}}}

\newcommand{\tr}{\textsf{Tr}}

\newcommand{\be}{\begin{equation}}
\newcommand{\ee}{\end{equation}}
\newcommand{\e}[1]{\mathrm{e}^{#1}}

\begin{document}
\title{Quantum Gravity at Low Energies}
\author{David Wallace\thanks{Department of Philosophy / Department of History and Philosophy of Science; email \texttt{david.wallace@pitt.edu}}}
\maketitle

\begin{abstract}
I provide a conceptually-focused presentation of `low-energy quantum gravity' (LEQG), the effective quantum field theory obtained from general relativity and which provides a well-defined theory of quantum gravity at energies well below the Planck scale. I emphasize the extent to which some such theory is required by the abundant observational evidence in astrophysics and cosmology for situations which require a simultaneous treatment of quantum-mechanical and gravitational effects, \emph{contra} the often-heard claim that all observed phenomena can be accounted for either by classical gravity or by non-gravitational quantum mechanics, and I give a detailed account of the way in which a treatment of the theory as fluctuations on a classical background emerges as an approximation to the underlying theory rather than being put in by hand. I discuss the search for a Planck-scale quantum-gravity theory from the perspective of LEQG and give an introduction to the Cosmological Constant problem as it arises within LEQG.
\end{abstract}

\section{Introduction}

What is a theory of `quantum gravity'? It is widely described\footnote{See, \egc, \citeN[p.33]{butterfieldishamchallenge}: ``by `quantum gravity' we mean any approach to the problem of combining (or in some way `reconciling') quantum theory with general relativity''; \citeN{rovelli-scholarpedia}:``Quantum Gravity is the name given to any theory that describes gravity in the regimes where quantum effects cannot be disregarded''; \citeN[p.262]{ricklesQGreview}:``Quantum gravity involves the unification of the principles of quantum theory and general relativity''.} as any physical theory which combines general relativity and quantum theory; or, alternatively, as any physical theory that has classical general relativity and quantum theory as limiting cases or partial approximations.

And here are some common claims about a quantum theory of gravity, understood in this sense:
\begin{enumerate}
\item We don't have one.\footnote{``[t]here is still no satisfactory theory'' \cite[p.33]{butterfieldishamchallenge}; ``an established theory of Quantum Gravity does not yet exist''\cite[p.265]{ricklesQGreview}.}
\item More specifically, we don't have an acceptable quantum theory of gravity in the sense that we have an acceptable quantum theory of the strong, weak or electromagnetic interactions.\footnote{``[C]urrently there is no quantum theory of gravity in the sense that there is, say, a quantum theory of gauge fields'' \cite[p.3]{callenderhuggettintroduction}.}
\item \emph{Empirically} we don't really need one, because --- outside exotica such as Hawking radiation, the interior of black holes, and the very early Universe --- all known empirical phenomena are describable either within quantum mechanics (more specifically: within the Standard Model of particle physics) or within classical General Relativity. So any case for a quantum theory of gravity will have to be based largely on theoretical considerations.\footnote{``there are currently no observations demanding a quantum gravitational theory'' \cite[p.3]{callenderhuggettintroduction}; ``there are no phenomena that can be identified unequivocally as the result of an interplay between general relativity and quantum theory'' \cite[p.36]{butterfieldishamchallenge}; ``If there is no empirical anomaly, then why bother with quantum gravity: why do we need such a theory?'' \cite[p.268]{ricklesQGreview}.}
\end{enumerate}
All of these claims are false. If `quantum gravity' indeed just means a theory that combines quantum mechanics with general relativity, then we have a fairly satisfactory theory of quantum gravity; indeed, we have a theory of quantum gravity that's acceptable in pretty much exactly the sense that we have an acceptable quantum theory of electrodynamics or of the strong or weak interactions. And there is abundant observational evidence for this theory, because there are an abundance of observations that cannot be explained in their entirety either by classical general relativity or by non-gravitational quantum mechanics, but which this theory describes admirably. Indeed, an important approximation regime for this theory --- often called `semiclassical gravity' --- is the (admittedly often tacit) foundation of most of contemporary astrophysics and cosmology. Other regimes are less directly tested but are still have applications in physical situations much more amenable to direct observation than the extremes usually associated with evidence for quantum gravity.

This theory is simply the quantum field theory (QFT) obtained when modern quantum-field-theoretic methods (specifically, path-integral formulations and the effective-field-theory approach) are applied to classical general relativity. It has its share of conceptual puzzles and problems, to be sure --- most seriously, it ceases to be well-defined when used to describe phenomena at extremely short distances, extremely high energies, and extremely high curvatures (more precisely, at lengths which are not large compared to the Planck length, $l_P=\sqrt{\hbar G/c^3}\simeq 10^{-35}\mathrm{m}$, and similarly for other magnitudes).  But this deficiency, far from being unique to the theory, is actually shared by the Standard Model and (arguably) by classical general relativity: both predict the appearences of extremes in which they can be expected to break down. 

In this paper I call it \emph{low-energy quantum gravity}, or LEQG for short. (Extant names are either too clunky (`effective-field-theory-general-relativity') or too ambiguous for my purposes (`general relativity', `quantum general relativity')). It is worth emphasizing that `low energy' here means `low energy relative to Planck scales', which is an undemanding constraint: the heart of a supernova, and the quark-gluon plasma of the early Universe, are both low-energy regimes. The purpose of this paper is to give a presentation of LEQG and of the observational evidence that supports it, and to consider how the broader project of quantum gravity should be understood in the light of LEQG. I intend the paper as a conceptual presentation complementing more calculationally-oriented discussions, and for the most part assume without proof (but with references) the mathematical results I use. (The main exception is the somewhat-explicit presentation I give of the background-field method in sections \ref{background-field-methods}--\ref{background-gravity}.)

The paper is structured as follows. In sections \ref{examples1}--\ref{examples2} I review a range of astrophysical and cosmological applications which require simultaneous consideration of quantum theory and gravity for their analysis. In section \ref{semiclassical} I present `semiclassical gravity', a hybrid theory where the metric field couples to the expected value of the stress-energy tensor, as a unifying framework for these applications, and argue that the observational evidence for `classical' general relativity mostly supports either semiclassical gravity or phenomenological applications of general relativity. In sections \ref{nr-fluctuations}--\ref{inflaton} I discuss the limitations of semiclassical gravity and the observational evidence --- both at laboratory scales and in cosmology --- that points beyond it, to a theory of quantized metric fluctuations.

In sections \ref{background-field-methods}--\ref{background-gravity} I present LEQG in general, stressing its non-perturbative and background-independent character and explaining how background-dependent descriptions arise from it as contingent approximations. In section \ref{LEQG-scope} I consider to what extent LEQG can be seen as empirically confirmed, and identify those regimes in which it is and is not supported by evidence. In section \ref{high-energies} I consider the breakdown of LEQG at Planckian scales and the question of how to seek a theory applicable at those scales. And in section \ref{ccp} I give an introduction to the notorious \emph{cosmological constant problem}, which arises in LEQG simply as a matter of calculation. Section \ref{conclusion} is the conclusion.

A note on level and rigor: I assume some background knowledge of general relativity and of quantum field theory in the path-integral formulation. I have aimed for about the level of mathematic rigor found in theoretical high-energy physics, which of course is lower than that found in pure mathematics, in some areas of general relativity, and in some parts of the foundational literature. For those who find results stated at this level unconvincing, I hope the paper will at least be illuminating as to how quantum gravity looks from the perspective of mainstream theoretical physics.

A note on sources: the particular presentation of LEQG I give here is novel in the minimal sense that I worked it out for my pedagogical purposes and have not straightforwardly followed any single source, but I claim no originality in the broader sense. Good general sources for effective-field-theory discussions of gravity, from a perspective somewhat complementary to my own and providing more on the technical details, are (\citeNP{burgess-eft}, Donoghue \citeyearNP{donoghue-eft-1997,donoghue-eft-2012}; \citeNP{kieferqg},ch.2).

Finally, a note on units: I take $\hbar=c=1$ except where explicitly noted otherwise, so that mass has the dimensions of inverse length, and so I talk interchangeably about physics at high energies and at short lengthscales. I do not take $G=1$ (in low-energy quantum gravity this is often unhelpful). Note that in these units $G$ is simply related to the Planck length, $G=\sqrt{l_P}$.

\section{Observing quantum gravity: the Newtonian regime}\label{examples1}

Take a rock, or any other solid object. Hold it in your hand, a few feet above the floor, and let go. Observe its steady downward acceleration followed by a sharp deceleration when it hits the floor.

Congratulations! You just carried out an experiment in quantum gravity. (Who said fundamental physics experiments had to be expensive?) The rock accelerated downwards because of gravity: if you ignore gravity and just treat the reference frame of the room you're standing in as inertial, you won't predict that acceleration. It stopped because it and the floor are made of non-interpenetrable solid matter, and to explain \emph{that} requires the Pauli exclusion principle (there is no classical microscopic theory of stable solid matter).

That experiment only required Newtonian gravity and quantum theory. To find an experiment that needs both \emph{general relativity} and quantum theory, just check the map app on your cellphone. Famously,\footnote{For details, see \cite{ashbyGPS}.} the GPS network requires general relativity for a full analysis; the processor in your phone needs quantum theory for a full analysis; the covariation of your physical position with your apparent position on the map cannot be fully analysed without both general relativity and quantum theory. Nor can the various classic experiments that use atomic clocks to measure general-relativistic time dilation.

This probably seems like cheating. In each case, there is a \emph{background} classical gravitational field against which quantum phenomena played out. One might imagine a division of labor: classical general relativity tells us the background gravitational field (or the spacetime metric, if you prefer) and only then do we apply quantum mechanics. But it at least makes the point that we have to be a bit careful saying what we mean when we say that ``there are no phenomena that can be identified unequivocally as the result of an interplay between general relativity and quantum theory'' \cite[p.36]{butterfieldishamchallenge}.

Here is an attempt to be more careful: for an experiment or observation to require a quantum-gravitational description, it must require a quantum-mechanical description of \emph{self-gravitating} matter. None of the experiments above fit this description: in each, a quantum-mechanical system (the rock, the satellite/cellphone system, the atomic clock) is moving in a gravitational field, but is not being treated as the source of a gravitational field.

But it is also easy to find situations which are quantum-gravitational in this more careful sense. Again: there is no classical microscopic account of solid matter. So a full physical description of the structure of Earth, or any other planet,\footnote{Even gas giants have solid cores; nor is this a contingent accident: self-gravitating gas clouds aren't stable, because of the gravothermal catastrophe (cf \cite[pp.567-572]{binneytremaine} and references therein).} requires quantum theory. But of course it also requires gravity: a physical analysis of the structure of a planet involves solving for its density and pressure as a function of its radius, taking as input (a) the equation of state of the matter forming the planet, which requires quantum theory to explain, and (b) Newton's law of gravity.

One might object that geologists do not in practice use quantum theory to calculate the equation of state of the Earth's matter: perhaps we should understand that equation of state as phenomenological rather than quantum-mechanical. Replying to that objection would take us deep into both the technical details of geology and the philosophical problems of reduction and emergence, so I will bypass it by considering the technically simpler case of stars.\footnote{The physics I discuss here is well-understood; (Kippenhahn~\emph{et al}~\citeyearNP{kippenhahn}) is a standard reference.} Stars like the Sun can be reasonably well modelled, at least in their broad-brush features, as balls of dilute gas or plasma, both of which are pretty well understood in microphysical terms (at least if we bracket philosophical problems in statistical mechanics). Now granted, the equations of state for plasmas and gases can be understood classically\footnote{Or at least, let's stipulate that they can; for some reasons for doubt, see \cite{wallacequantumstatmech}.} --- but there is no classical account available of nuclear fusion, the process by which a star's temperature is maintained against the constant loss of energy through radiation. Nor is there a classical account available of how photons diffuse from the stellar core to the star's atmosphere and so are visible as sunlight. A proper microscopic understanding of fusion requires a quantum-mechanical account of the nucleus and of the strong interaction; a proper microscopic understanding of stellar opacity requires quantum electrodynamics. So a full account of a star requires both quantum theory and gravity.

If we consider stars smaller than the sun, quantum mechanics enters even into their equations of state. In smaller stars the core is maintained by \emph{electron degeneracy pressure}: in semiclassical terms, the Pauli exclusion principle permits only one electron
of each spin in a phase-space region of size $\hbar^3$, so as the spatial density of electrons becomes larger they are forced into higher-and-higher-momentum states in order to conform to the exclusion principle. (This is well-understood physics: the semiclassical and verbal gloss I give is replaceable by sharp mathematics.) The same phenomenon occurs in white dwarf stars, the corpses of stars like the Sun which have exhausted their fusion fuel. In either case, the structure of the star is explained, and indeed calculated quantitatively and checked against observational data, using a marriage of quantum theory and Newtonian gravity.

\section{Observing quantum gravity: relativistic matter and strong gravity}\label{examples2}

The examples I have given so far are largely nonrelativistic: we need to consider relativistic physics to handle fusion reactions, and we need to consider electromagnetic effects to understand heat flow around stars, but the short-term (hydrostatic) descriptions of planets, main-sequence stars, and white dwarfs rely just on nonrelativistic quantum mechanics and on Newtonian gravity. But it is not hard to find examples in astrophysics where Newtonian gravity is insufficient and relativistic effects must be considered, but where quantum theory remains essential.

The first class of examples arise where it matters that Einsteinian gravity couples to the full stress-energy tensor of matter, and not just to its rest mass. For instance, in cosmology we identify `radiation' and `baryonic matter' components of the overall matter distribution (along with the more mysterious `dark matter' and `dark energy' components). As a capsule summary of their physics:
\begin{itemize}
\item The radiation can basically be treated as black-body; its energy density is proportional to the fourth power of its temperature.
\item The baryonic matter can be treated initially as a relativistic plasma, with appropriate equation of state; in due course it cools down to a nonrelativistic gas, at which point its energy density is effectively independent of its temperature, and inversely proportional to the Universe's effective volume.
\item Initially the radiation and matter are in thermal equilibrium with one another and so at the same temperature. Eventually (specifically, when protons and electrons combine to form hydrogen, making the baryonic matter transparent) the interaction between the two drops off too rapidly to maintain equilibrium and they continue to evolve separately.
\item After this stage, the temperature of the radiation is inversely proportional to the universe's effective radius, and so the energy density of the radiation falls off more rapidly than that of the baryonic matter, and the latter dominates the former gravitationally.
\end{itemize}
This is standard, well-understood cosmology\footnote{See \cite{weinbergcosmology} or any other standard text for details.} --- but it requires both quantum theory and general relativity to understand. The energy density of black-body radiation is, famously, a quantum-mechanical result that cannot be reproduced classically --- indeed, exactly this problem led Planck to the idea of the light quantum. Similarly, the physics of how the radiation and matter move out of equilibrium is quantum-mechanical, turning on the interaction of photons first with plasma and then with atomic hydrogen. But that energy density is not a nonrelativistic `mass', and to properly couple it to gravity requires general relativity. Ordinary cosmology is an application of relativistic quantum gravity.

The second class of examples arise when the strong-gravity regime of general relativity comes to the fore. For instance, consider neutron stars, white dwarfs' more extreme cousins. In a neutron star, gravitation-induced pressure is so high that protons and electrons are induced to combine into neutrons, and the resultant neutron fluid then supports itself by neutron degeneracy pressure. The basic physics is similar to white dwarfs, but the much higher densities involved mean that a general-relativistic description of a neutron star's gravity is necessary to give an accurate analysis of its structure. The physics of neutron matter is also more complicated by far than that of the degenerate matter in white dwarfs; a full understanding remains elusive, but what understanding we have relies on a combination of finite-temperature quantum field theory\footnote{In a strict sense this is not a \emph{relativistic} application of quantum field theory --- the `finite temperatures' in question are much lower than the neutron rest mass --- but it certainly lies outside the domain of applicability of nonrelativistic particle mechanics.} and general relativity. (For details, see (Keller~\emph{et al}~\citeyearNP{kellerneutronmatter}) and references therein.)

For an example which combines both relativistic features of matter and the strong-gravity regime of general relativity, consider gravitational supernovae, the catastrophic explosion of (some) large stars when they exhaust their nuclear fuel. In outline, sufficiently large stars build up a core of degenerate iron (iron cannot release further energy through nuclear fusion). When this core exceeds a certain mass it becomes unstable due to relativistic effects and collapses very rapidly in on itself to form a neutron star. The remaining matter falls onto the neutron core and `bounces' (in a fashion which remains incompletely understood), dispersing into a large cloud around the neutron star, accompanied by a catastrophic release of energy, mostly in the form of neutrinos. The core of a supernova is an extreme physical environment. As David Arnett puts it in his widely-cited monograph on supernovae and nucleosynthesis,
\begin{quote}
The problem abounds with the extreme and the exotic. Neutrinos, which eluded direct detection for decades because of the extreme weakness of their interaction, are trapped in the collapse. The collapse is so fast that a mass, more than half that of the sun, shrinks in on itself at velocities approaching the speed of light. Einstein's general relativity theory becomes necessary to describe the event, not merely a subtle correction to the gravitational theory of Newton. Perhaps nowhere else in the Universe does it matter so much whether neutrinos obey Pauli's exclusion principle.
 \cite[p.381]{arnett}
\end{quote}
(For further details of the physics here, see (Arnett \emph{ibid}.) and references therein.)

To conclude this part of the discussion: it is flatly wrong that contemporary physics provides few or no situations in which gravity and quantum theory must be considered in parallel. Astrophysics and cosmology routinely give us situations in which quantum-mechanical and self-gravitation effects must be considered in parallel, and many of these situations involve core features of relativistic gravitation and relativistic quantum theory.

\section{Semiclassical gravity and the observational evidence for general relativity}\label{semiclassical}

If astrophysics and cosmology are routinely using both quantum-mechanics and gravitational physics in the same problems, it ought to be possible to say --- at least in a rough-and-ready way --- what the principles are that govern their combination. And indeed, I think it is fairly straightforward to read off from these applications how this goes. In each case, quantum mechanics is used to give us appropriate dynamical equations for bulk matter, using the usual rules of quantum statistical mechanics: work out the expected values of quantities like fusion rates, pressures, and energy densities, and then treat these as the actual values of those quantities under the assumption that quantum fluctuations in their values are negligible (and if necessary check that assumption by calculating the variance in those values using the quantum formalism, though in most cases it is obvious on physical grounds that they are negligible). Normally the quantum calculations are carried out on flat spacetime, on the assumption that the dynamically relevant scales for matter are much smaller than the relevant spacetime curvature scales, and translated to general curved spacetimes via the equivalence principle. Then the matter equations thus obtained are combined with Einstein's field equations and solved.

This approach, where quantum-mechanical matter is analysed on a classical spacetime and the two are coupled by putting the expected value of the stress-energy tensor into the Einstein field equations ---
\be\label{semiclassical-eqn}
G_{\mu\nu}[g] + \Lambda_0 g_{\mu\nu} = 8 \pi \langle T_{\mu\nu}\rangle_{\ket{\psi}}
\ee
--- is generally referred to as \emph{semiclassical gravity}. It has been reasonably widely discussed in the quantum-gravity literature (see, \egc, \citeNP[pp.15-21]{kieferqg}, \citeNP[p.8]{wallgsl}, and references therein), but it is not often recognized in that literature that  --- as the examples of sections \ref{examples1}--\ref{examples2} demonstrate --- it has widespread applications in extant physics, encompassing much of contemporary astrophysics and cosmology.

Indeed, we can go further: in an important sense, semiclassical gravity is more strongly supported by observational evidence than `classical' general relativity. In the abstract, modelling a system via general relativity requires:
\begin{enumerate}
\item A class of models $\langle \mc{M}, g, \phi \rangle$, where $\mc{M}$ is a four-dimensional manifold, $g$ is a Lorentzian metric on that manifold, and $\phi$ is a placeholder for whatever mathematical objects on the manifold represent the matter fields.
\item Some kind of dynamical equations for the matter fields $\phi$, expressed on the background metric $g$.
\item A dynamical equation of the form
\be
G_{\mu\nu}[g] + \Lambda g_{\mu\nu} = 8 \pi \mc{T}[\phi,g]
\ee
where $\mc{T}[\phi,g]$ is some kind of expression that determines the right-hand side of the equation in terms of the matter fields and the metric.
\end{enumerate}
Together, this (at least formally) forms a closed dynamical system for $g$ and $\phi$, which we can then solve to determine which of the triples $\langle \mc{M}, g, \phi \rangle$ represent real dynamical possibilities.

There are essentially four concrete versions of this schema that are discussed in physics. The first is \emph{vacuum general relativity}, where there are no matter fields at all and the right hand side of the Einstein field equation is set to zero. The second is \emph{phenomenological general relativity}, which eschews any microphysical account of matter and uses `dust', `perfect fluid', `point particle' and similar models. The third is semiclassical gravity, which we have already discussed. And the fourth is \emph{classical general relativity}, where a classical matter field like the Klein-Gordon field, Dirac field or electromagnetic field is coupled to gravity, usually using an action principle to determine both the dynamical equations for matter and the stress-energy tensor.

The first three of these find fairly widespread application in astrophysics. Vacuum general relativity is applied to the study of black holes and gravity waves; the calculations underpinning the LIGO gravity wave observatory's fit to predictions rely on it (cf. \cite{ligoreview} and references therein).
Phenomenological general relativity (albeit often in a Newtonian approximation) is commonly applied in various pieces of galactic-dynamics and cosmology, where the details of matter can be abstracted away, and also in solar-system physics, where Birkhoff's theorem\footnote{See, \egc, \cite[p.125]{waldrelativitybook} or \cite[p.197]{carrollgr}.} allows us to disregard the details of the Sun's structure given that it is spherically symmetric. And we have seen that semiclassical gravity is necessary whenever we need a non-phenomenological description of quantum-mechanical matter. In practice the line between phenomenological general relativity and semiclassical gravity is blurry: it is often a matter of degree in astrophysics to what extent the dynamical equations of matter are derived bottom-up and to what extent they are modelled phenomenologically.

Classical general relativity, however --- despite its undoubted mathematical beauty --- has no applications at all to observational data (except insofar as we treat it as including vacuum general relativity as a special case). To the best of my knowledge, all observationally-relevant treatments of matter coupled to general relativity use some mix of phenomenological and quantum-mechanical treatments: there are no observationally-relevant cases where a classical field, in the usual sense of that term, is coupled to gravity. Nor should this be surprising: the only classical fields seem to be\footnote{There are some field-theoretic models of dark matter (see \cite{scalarfielddarkmatter-review} for a review), so it would be premature to absolutely rule out a third classical field.} gravity itself and the electromagnetic field, with the other quantum fields of the Standard Model manifesting in the classical regime, if at all, as particles. And while in principle a classical electromagnetic field could generate a gravitational effect, in practice this does not seem to occur in astrophysically-relevant situations.\footnote{The nearest I can imagine to an exception might be the enormously strong magnetic fields generated by neutron stars and by black hole accretion disks. But even there, a self-contained account would require a quantum-mechanical and/or phenomenological description of the generation of the field, and could not be analyzed just using the classical theory of electromagnetism coupled to gravity. A black hole with a significant electric charge \emph{would} be modelable in that classical theory, but such objects do not appear to occur in nature, for physically obvious reasons.} (Black-body electromagnetic radiation self-gravitates non-trivially in some circumstances, but as we have already noted, this requires a quantum-mechanical treatment.)

So: what is `general relativity', if we mean the theory which is actually used and applied in observational physics? It is not a self-contained classical theory that could be thought of as fundamental were it not for the pesky quantum effects in other regimes: it is in practice a higher-level effective theory, in which matter is treated either phenomenologically or quantum-mechanically through the methods of semiclassical gravity.\footnote{It is also non-fundamental for another reason lying somewhat outside the main scope of this paper: in at least a large fraction of applications of general relativity, including almost all applications at very large scales, the metric tensor $g$ is not the `true' metric (whatever that means) but a highly coarse-grained approximation to that metric, smoothing out metric inhomogeneities over macroscopically large regions. Cosmology often smooths out even the structure of galaxies, let alone the structure of the matter within them. See \cite{wiltshirewhatisdust} and references therein for further discussion of this point.}

\section{The limitations of semiclassical gravity}\label{nr-fluctuations}

Semiclassical equations like (\ref{semiclassical-eqn}) are very common in physics. Known variously as `Vlasov equations', `mean-field approximations', and `Hartree(-Fock) approximations', they have extremely wide applicability, from nuclear and atomic physics through to galactic dynamics. The Hartree approximation to nonrelativistic quantum electrostatics, for instance ---
\be
i \hbar \pbp{\psi}{t}(x,t) =  - \frac{\hbar^2}{2m}\nabla^2 \psi(x,t) - e V(|\psi|)(x,t)\psi(x,t)
\ee
where 
\be
V(|\psi|)(x,t)=\int \dr{x'}^3 \frac{e|\psi(x',t)|^2}{|x-x'|}
\ee
--- describes electrons moving in the mean electric field created by those same electrons, and is the starting point for much of atomic physics and quantum chemistry.\footnote{Strictly speaking, the equation used is the Hartree-Fock equation, which takes better account of the antisymmetrisation properties of a multi-electron wavefunction.} But normally these equations are taken as approximations to the underlying dynamics, approximations which hold only on the assumption that multiparticle correlations, or their field-theoretic equivalents, are low. (The Hartree equation, for instance, is derivable\footnote{See, \egc, \cite{bonitz}.} as a first-order approximation to the $N$-particle Schr\"{o}dinger equation under the approximation that the quantum state remains approximately in product form).
And there is a clear physical reason why these approximations ought to fail in the presence of strong correlations: if the quantum state of $N$ electrons is a superposition of the electrons at $X$ and the electrons at $Y$, an $N+1$th electron interacting with the first $N$ would be expected to end up entangled with them, not feeling the electric field as if its source was at $(X+Y)/2.$ Adopting the language of decoherent histories,\footnote{(Gell-Mann and Hartle~\citeyearNP{gellmannhartle,gellmannhartle93}; Griffiths~\citeyearNP{griffiths,Griffiths1993}; Omnes~\citeyearNP{omnes,omnes92}); see also reviews by \citeN{halliwellreview} and \citeN[ch.3]{wallacebook}. } an electron in a history $h$ ought to feel the electric field of the other electrons at their positions as described in $h$, not their expected position averaged over all histories.

It is then natural to expect that semiclassical gravity is likewise an approximation to some more fully quantum theory, and that the equations of semiclassical gravity would fail under the same circumstances that other semiclassical methods fail: that is, when the quantum state is a superposition of substantially different mass distributions (something that is not the case\footnote{There is a caveat here: given unitary dynamics, the \emph{universal} quantum state (insofar as such a thing makes sense) is certainly in a superposition of different mass distributions. I discuss this below.} in any of the astrophysical and cosmological examples we have considered so far). And it is fairly straightforward to test this empirically: simply use a quantum-mechanically random process (a spin measurement, say) to place a mass large enough to generate a detectable gravitational field in one of two macroscopically distinct locations in the lab; then measure the gravitational force. If semiclassical gravity fully describes quantum gravity in this regime, the force should be detected as directed towards an empty location intermediate between the two locations; if semiclassical gravity is a mean-field approximation to something else, we would expect the force to be detected as directed towards the observed location of the mass. This experiment has actually been done \cite{Page1981}, demonstrating (perhaps unsurprisingly) that semiclassical gravity indeed fails in this situation. 

Similarly, considerations of the long-term chaotic behavior of the solar system, in which one would expect the locations of the planets to become macroscopically superposed on megayear timescales \cite{zurekpazplanetary}, might be said to provide an observational version of the Page-Geilker experiment, as Page and Geilker themselves note (\emph{ibid}., p.980). On sufficiently long timescales, no unitarily-evolving quantum state will remain macroscopically definite.\footnote{For that reason, at least on interpretations which treat the quantum state as observer-independent, the empirical evidence for semiclassical gravity really supports what we might call \emph{branch-relative semiclassical gravity}: semiclassical gravity applied not to the universal state but to the branch-relative state.}

The Page-Geilker experiment has received some pushback in the philosophy literature, notably by Mattingly~(\citeyearNP{mattinglynecessary,mattinglymongrel}). Mattingly's objection is that Page and Geilker assume the Everett interpretation of quantum mechanics (they do indeed mention it explicitly in their paper) and that their experiment does not have any implications for semiclassical gravity without this assumption. 
\begin{quote}
I have no solution to the quantum measurement problem. Nor do I find wholly satisfying any of the extant proposals for solving it. Yet I cannot worry overmuch about a proposal for a gravitation theory merely because it fails to solve the problem, or rather because it undermines one or two of the proposed solutions. \cite[pp.330-331]{mattinglynecessary}.
\end{quote}
But this understates the scope of the Page-Geilker result (and of similar, more theoretical arguments; cf \cite[pp.15--22]{kieferqg} and references therein). The argument I sketched above does not rely on any \emph{interpretative} assumptions, but just on the \emph{mathematical} assumption that the quantum state of the lab in which the experiment is conducted evolves unitarily (or, more precisely, according to the modified version of unitarity that incorporates the semiclassical gravitational field). The Everett interpretation makes this assumption, but it is not alone in doing so: the various consistent- and decoherent-history approaches to quantum mechanics all assume unitary dynamics, as do the various hidden-variable and modal `interpretations'.\footnote{Hidden-variable theories, to be sure, also allow for the possibility that the gravitational field is sensitive to the hidden variables and not (or not just) to the physical features encoded in the quantum state; see \cite[section 8]{pinto-neto-struyve-gravity} for further development of this idea. However, theories of this kind take us beyond semiclassical gravity, in which the right-hand-side of the Einstein equation is a quantum expectation, independent of the hidden variables. (For that reason they also face a potentially severe observational challenge in relativistic regimes, where a quantum calculation of the stress-energy tensor is used in astrophysics and cosmology.) } Even pragmatist or instrumentalist approaches such as QBism (Fuchs \emph{et al},~\citeyearNP{fuchsmerminschack}), Healey's pragmatism~(Healey \citeyearNP{healeypragmatism,healeypragmatismbook}), Fuchs' and Peres' operationalism~\cite{fuchsperes}, or most versions of the Copenhagen interpretation at least permit the lab to be analyzed unitarily, since they all permit Wigner's-friend scenarios where an external observer considers the lab (experimenters and all) as a single quantum system. This same move is permitted on at least some interpretations of Dirac and von Neumann's `measurement-collapse' approach, since that approach is often taken to allow us to put the transition between unitary and non-unitary dynamics (von Neumann's `cut') as close to the observer as we wish. It is permitted, too, under Rovelli's relationalist interpretation~\cite{rovelli-relational}. Of the various interpretations of quantum mechanics, and proposals for modifications of quantum mechanics, currently extant in the philosophy and physics literature, only dynamical-collapse theories clearly prohibit a unitary analysis of the Page-Geilker experiment. And while the suggestion that dynamical collapse might have some connection to gravity is certainly intriguing, and has been the source of interesting theoretical speculations and experimental proposals (to which I return in section \ref{LEQG-scope}), at this point we are well outside anything that might be called an `interpretation' of quantum mechanics. The mainstream use of quantum theory in modern physics (cf Wallace~\citeyearNP{wallaceorthodoxy,wallacecosmology}) assumes unitary quantum dynamics, and that is all that is required for Page-Geilker-type experiments to falsify semiclassical gravity. 

Nonetheless, the Page-Geilker experiment --- and Zurek and Paz's theoretical models --- have a pretty `foundational' feel to them, concerning as they do macroscopic superpositions which are implied by the equations but not in any direct sense observable. There is, however, an important class of applications of quantum gravity in mainstream cosmology where the presence of quantum fluctuations in the gravitational field, and the responsiveness of matter to those fluctuations and not just to their expected value, has direct observational consequences. These applications --- to the quantum origins of inhomogeneity in the microwave background radiation and in the cosmological-scale distribution of matter --- will be of interest both because they further mark the limitations of semiclassical gravity and because they provide further examples of the direct use of quantm gravity in contemporary physics.

\section{Quantum fluctuations in the inflaton field}\label{inflaton}

A key feature of gravitational physics, owing its origin to the universally-attractive nature of gravity, is that small fluctuations from isotropy and homogeneity in the matter distribution will grow exponentially in size over time (rather than be damped back to uniformity as in the case of, say, a non-self-gravitating dilute gas or electromagnetic plasma). As such, even very small fluctuations away from uniformity in the early Universe can suffice to generate the highly non-uniform cosmos we observe today, in which typical matter densities in stars exceed the average matter density of the Universe by around thirty orders of magnitude. When in 1992 the COBE satellite detected relative fluctuations of $~10^{-5}$ in the observed cosmological microwave background (CMB), this began the modern era of understanding in quantitative detail exactly how these small fluctuations indeed seed the large-scale structure we observe today, and arguably can be seen as the beginning of the current golden age of observational cosmology. (Here and for the rest of this section, see \cite{weinbergcosmology} for details and references.)

The \emph{origin} of these fluctuations is another matter. As long as we assume a classical, deterministic picture of cosmology, small fluctuations in the CMB must have their origin in still-smaller --- but still non-zero --- fluctuations in the still-earlier Universe. But the dominant view in modern cosmology is that ultimately we should \emph{not} assume that `classical, deterministic picture', and instead should look to understand these primordial fluctuations as the classical remnants of \emph{quantum fluctuations} in the early universe, even given an isotropic and homogeneous quantum state for the early universe -- and that concrete quantum theories of those fluctuations can be used to calculate a predicted spectrum of fluctuations, compared to observation, and thus tested. The dominant such theory, \emph{inflation}, has indeed been so tested, and while the `test' in question is scarcely precise, inflation does indeed correctly predict the form of the CMB spectrum. This connection of inflation to the CMB and thus, indirectly, to large-scale structure formation has probably now replaced older, more foundational, motivations as the main reason for its popularity amongst cosmologists (see \cite{smeenk-testinginflation} for more discussion).

It should be clear on physical grounds that any such quantum analysis of fluctuations cannot assume semiclassical gravity. If we begin with a homogeneous and isotropic quantum state describing matter on a homgeneous and isotropic background, and allow it to evolve semiclassically, its expectation values will remain homogeneous and isotropic. From the quantum point of view, structure formation is a correlational phenomenon invisible to mean-field methods. And indeed, direct examination of the sorts of calculations used in this field confirms that they are not made within semiclassical gravity. Instead, the usual approach is to split the metric field $g$ into a sum $g=g_0+h$, where $g_0$ is homogeneous and isotropic and $h$ thus represents fluctuations around $g_0$, and then to quantize $h$ and the matter field (say, an inflaton field $\phi$) as quantum fields interacting with one another on a background spacetime with metric $g_0$. Fluctuation information is then contained in expectation values like $\langle \op{h}(x)\op{h}(y)\rangle$, which, given decoherence, can be interpreted as encoding correlations between fluctuations at different points.

In  the most straightforward of these analyses, the radiational degrees of freedom of $h$ (i.e., the gravitons) are assumed neglectable, and so $h$ is basically a functional of the matter field, representing gravitational interactions between matter fluctuations. In more detailed analyses (e.g., Bhattacharya~\emph{et al},~\citeyearNP{bhattacharya-graviton-background}) gravitons are also present. But even when gravitons are frozen out, this is fairly clearly a fully quantum theory of gravity, with the gravitational field being a superposition of different classical values, and entangled with the matter field. 

It is fair to say that the physics here is significantly more speculative and significantly less observationally constrained than the very mature astrophysics and cosmology that underlies applications of semiclassical gravity. But it is also fair to say that it is much \emph{less} speculative, and much \emph{more} observationally constrained, than proposals for Planck-scale quantum gravity. Inflationary models do make predictions about the quantitative form of the CMB fluctuations; those predictions are non-trivial; they have so far been confirmed. And active efforts continue to further constrain inflationary models with observational data: notably, the BICEP and Keck arrays at the South Pole\footnote{See Ada~\emph{et al}~\citeyear{bicepreview} and references therein.} aim to measure polarization in the CMB, which is an indirect measure of the graviton background in the primordial Universe.

\section{Background field methods}\label{background-field-methods}

The applications I have so far discussed demonstrate fairly widespread applications of quantum gravity in contemporary astrophysics and cosmology, and comparably widespread exposure of those applications to observational test. But at first sight they also paint a rather messy disunified picture of quantum gravity at low energies. Semiclassical quantum gravity has a fully dynamical interaction between quantum matter and spacetime metric, but treats the latter as fully classical, and is for that reason (as we have seen) not satisfactory except as some sort of approximation. But it is not obviously any kind of approximation to the essentially perturbative approach to quantum gravity used in inflation, in which a classical background spacetime plays a central role and in which quantization is apparently only defined relative to that background.

This perception is misleading. The methods of modern quantum field theory --- specifically, the `background field method', developed by Goldstone, Salam and Weinberg in particle physics, and independently by de Witt in the context of quantum gravity (see \citeN[ch.16]{weinbergqft2} for a review and for original sources) provide an elegant way to see both semiclassical methods and quantize-on-a-background methods as different approximation schemes for the same underlying theory.  In this section I give a brief outline of the background field method in the simpler context of scalar field theory; section \ref{background-gravity} applies the method to gravity.\footnote{For calculational uses it is more convenient to develop the background field method using the powerful and elegant \emph{quantum effective action} (see \cite[ch.11]{peskinschroeder} for an introduction and (e.g.) \citeN[ch.16]{weinbergqft2} or \citeN[ch.2]{banksQFT} for more advanced treatments). But this is overkill for my purposes; here I adopt a more heuristic approach.}

In contemporary QFT, the theory is normally specified by means of a \emph{path integral}, which for scalar field theory has the form
\be\label{pathintegral}
\bk{\phi_f;t_f}{\phi_i;t_i}= \int_{\phi(t_i)=\phi_i}^{\phi(t_f)=\phi_f}\mc{D}\phi\,\e{i S[\phi]}
\ee
where 
\begin{itemize}
\item $\ket{\phi_i;t_i}$ is a quantum state in the Heisenberg representation of the theory, defined by $\op{\phi}(\vctr{x},t_i)\ket{\phi_i;t_i}=\phi_i(\vctr{x},t_i)\ket{\phi_i;t_i}$, and similarly for $\ket{\phi_f;t_f}$.
\item  $S[\phi]$, the \emph{action} for the scalar field theory, is a quasi-local functional of the classical scalar field $\phi$, which we assume takes the general form
\be\label{action}
S[\phi] = \int \dr{x^4} \left(\frac{1}{2}\partial^\mu\phi(x)\partial_\mu\phi(x) - V(\phi(x)) \right).
\ee
\end{itemize}
That is, the amplitude for transitions between configuration $\phi_i$ at time $t_i$, and configuration $\phi_f$ at time $t_f$, is expressed as a path integral over all histories that connect $\phi_i$ to $\phi_f$, with each history weighted by a phase factor determined by the action of the history. As stated this is all rather formal: the path integrals here are divergent and the theory must be regularized through the addition of some high-energy cutoff in order for the path integrals to be well-defined; I postpone consideration of these issues to section \ref{ccp}. 

Suppose we now fix a quantum state $\ket{\mathrm{in}}$, which we choose to encode a known state of the system at some early time. The physical features of this system at all times are encoded in expectation values like
\be
\langle F[\op{\phi}]\rangle_{\mathrm{in}} \equiv \matel{\mathrm{in}}{\op{\phi}(x_1)\cdots \op{\phi}(x_n)}{\mathrm{in}}
\ee
 and these can be expressed in terms of an appropriately manipulated path-integral: schematically, as
\be\label{exp-path-integral}
\langle F[\op{\phi}]\rangle_{\mathrm{in}} \propto  \int_{\mathrm{in}}=\mc{D}\phi F[\phi]\e{i S[\phi]}
\ee
where $`in'$ schematically indicates the boundary conditions on the integral required to enforce $\ket{in}$ as our choice of quantum state. The most common choices of state are the vacuum state $\ket{\Omega}$ and (through a generalisation to mixed states) the thermal state $\op{\rho}_\beta=\exp(-\beta \op{H})/\tr \exp(-\beta \op{H})$ which describes a system at equilibrium at temperature $T=1/\mathrm{k}_B\beta$, but we can in principle choose any state we wish, including states which are not eigenstates of energy (or mixtures of eigenstates) and so describe time-dependent phenomena. (In this latter case, the integral is the so-called `in-in' or `Schwinger-Keldysh' path integral, where the integral is taken on a double contour in time that extends into the far future and then back again, effectively doubling the fields present; for the details, see (e.g.) \citeN[section 3.1]{bergesqftreview} or \citeN[ch.6]{calzettahubook}.) From here on, I will usually omit the `in' subscripts, taking the specification of the state as tacit.

If the expectation value  $\phi_0(x)=\matel{\mathrm{in}}{\op{\phi}(x)}{\mathrm{in}}$ is zero, and if a number of other technical conditions hold, then we can evaluate these path integrals perturbatively and get increasingly good approximations for them. In these cases, we can normally (at least approximately, at least on short timescales) interpret the QFT in the standard way as a theory of spin-0 particles, and derive the masses, decay rates, and scattering cross-sections for those particles from the path integral. However, in many cases of interest $\phi_0$ is not zero, and these perturbative methods cannot straightforwardly be applied; this is the context for the background field method.

To begin with, suppose we carry out an infinitesimal change of variables in the path integral: $\phi(x)\rightarrow \phi(x)+\epsilon(x)$. The value of an integral is invariant under any such change, so
\begin{eqnarray}
0 = \int \mc{D}\phi \e{i S[\phi+\epsilon]} - \int \mc{D}\phi\e{i S[\phi]} &=&\int\mc{D}\phi \int \dr{x^4} i \vbv{S}{\phi(x)}\epsilon(x)\e{i S[\phi]}\nonumber \\
&=& i \int \dr{x^4} \epsilon(x)\int \mc{D}\phi\vbv{S}{\phi(x)}\e{i S[\phi]} \nonumber \\
&\propto &  i \int \dr{x^4} \epsilon(x) \left\langle \vbv{S}{\op{\phi}(x)}\right\rangle 
\end{eqnarray}
so that 
\be\label{expected-action}
\left\langle \vbv{S}{\op{\phi}(x)}\right\rangle=0.
\ee
(Since the variational derivative of $S$ is just the Euler-Lagrange equation, \ie the classical field equation for the action $S$, this can be thought of as the path-integral version of Ehrenfest's theorem.)

Now we make a \emph{finite} change of variables,
\be\label{finite-change-of-variables}
\chi(x)=\phi(x)-\phi_0(x).
\ee
(Recall that $\phi_0$ is the expectation value of the field on our specified quantum state $\ket{\mathrm{in}}$.)
We can now rewrite the action as
\be
S[\phi_0+\chi] = S[\phi_0] +\left.\vbv{S}{\phi}\right|_{\phi_0} \cdot \chi +\Sigma_{\phi_0}[\chi]
\ee
where in the second term on the right I use the abbreviated notation
\be
\left.\vbv{S}{\phi}\right|_{\phi_0} \cdot \chi \equiv \int \dr{x^4} \left.\vbv{S}{\phi(x)}\right|_{\phi_0} \chi(x).
\ee
Plugging this into (\ref{expected-action}), we obtain
\be
0 = \left.\vbv{S}{\phi}\right|_{\phi_0} + \left.\vbv{}{\phi}\right|_{\phi_0}\cdot  \left.\vbv{S}{\phi}\right|_{\phi_0} \langle \chi(x)\rangle + \left\langle \left.\vbv{\Sigma_\phi}{\phi}\right|_{\phi_0}\right\rangle. 
\ee
The second of these terms vanishes exactly, because by construction the expectation value of $\op{\chi}$ is zero, so that this simplifies to
\be \label{semiclassical-scalar}
0 = \left.\vbv{S}{\phi}\right|_{\phi_0}  + \left\langle \left.\vbv{\Sigma_\phi}{\phi}\right|_{\phi_0}\right\rangle. 
\ee
This is the sum of the classical Euler-Lagrange equation for the scalar field, and a quantum expectation value that can be expressed as a path-integral,
\be
\left\langle \left.\vbv{\Sigma_\phi}{\phi}\right|_{\phi_0}\right\rangle \propto \int \mc{D}\chi \left.\vbv{\Sigma_\phi[\chi]}{\phi}\right|_{\phi_0}\exp i \left(\chi\cdot  \left.\Sigma_{\phi_0}[\chi] + \vbv{S}{\phi}\right|_{\phi_0}  \right)
\ee
where I have just used the change of variables (\ref{finite-change-of-variables}) in (\ref{exp-path-integral}) and absorbed a constant phase factor into the constant of proportionality.

We have now reexpressed the original quantum field theory as a theory of two fields: a classical scalar field, and a quantum field whose action 
\be
S'_{\phi_0}[\chi]=\Sigma_{\phi_0}[\chi]+\chi\cdot  \left.\vbv{S}{\phi}\right|_{\phi_0}  
\ee
depends on that classical field. (The second term in this expression can be thought of as encoding quantum corrections to the classical expectation value and can largely be ignored in calculations, serving only to cancel certain diagrams in a perturbative expansion; it is the first term $\Sigma_{\phi_0}$ that really defines the interesting features of the quantum field.) The dynamics of the two fields have the classic form of a self-consistency problem: the quantum field evolves against a background defined by the classical field; the classical field responds to the quantum field via equation (\ref{semiclassical-scalar}). The expectation value of this new quantum field is zero by construction, and so we can hope (though must check explicitly) that we have a well-behaved perturbative description of that field. 

The best-known uses of the background field method are in spontaneous symmetry breaking (SSB), where (for the real field we are presently considering) we begin  with some potential like
\be
V(\phi)=- \frac{1}{2}\mu^2 \phi^2 + \frac{\lambda}{4!}\phi^4
\ee
for the action (\ref{action}). V has a maximum at $\phi=0$ and minima at $\phi=\pm (6/\lambda)^{1/2} \mu$; in elementary treatments of SSB  we pick one of these minima and quantize around it (much as we considered quantizing around a particular solution of the Einstein equations in section \ref{inflaton}). But the background-field method lets us give a fully quantum treatment, where we extremize not the classical action but the quantum effective action. The latter is dominated by the classical action but has quantum corrections which can be thought of as polarisations of the vacuum. These do not change the qualitative form of the function\footnote{This need not be the case. There are well-known examples (notably the Coleman-Weinberg model~\cite{colemanweinberg} where the classical action has a minimum with unbroken symmetry and the symmetry breaking has an entirely quantum origin.}, so that its extrema are homogeneous fields equal to 0 (a maximum) or $\pm \alpha$ for some nonzero $\alpha$ (a minimum). Calculating the two-point functions for the quantum fields defined around these extrema reveals that the maximum corresponds to a highly unstable quantum state and so does not describe a genuine vacuum, but that the minima are the expectation values of well-behaved vacuum states. The existence of multiple extrema tells us that the theory has multiple vacuum sectors, with the actual vacuum fixed mathematically by the boundary conditions, physically by contingent facts about the field-theoretic system being studied.
And crucially, we have obtained these sectors not by applying different quantization procedures for fluctuations around a background, but by defining the quantum theory once and for all by the path integral (\ref{pathintegral}) and then exploring its space of solutions.

If we analyze SSB for a finite-termperature state, the expectation value will now contain thermal contributions as well as vacuum polarization. These generally have the effect of decreasing the expectation value $\phi_0$, until for sufficiently high temperature $\Gamma$ ceases to have symmetry-breaking minima and the symmetry is restored (in particle-physics contexts, this is expected to happen in the early universe). 

A more complicated application of the background-field method is to situations where the quantum system starts in a strongly time-dependent state (\iec, not a vacuum state, a thermal state, or a state describable as a small perturbation of either). As already noted, this requires the more-complicated `in-in' path integral, which somewhat complicates the technical details of the background-field formalism. But the basic structure remains: we can use equation (\ref{semiclassical-scalar}) to express the dynamics of the system as a coupled interaction 
of a classical field and the quantized fluctuations around that field. The underlying theory remains fully quantum-mechanical, and this particular decomposition into background and fluctuation is solution-relative.

As a final observation before we turn to gravity: \emph{formally} the semiclassical equation (\ref{semiclassical-scalar}) is exact. But it gets its practical power from the ability to calculate a good approximation to the expectation-value term, and this in turn requires (among other things) that the strength of the fluctuations (expressed by the two-point function $\langle \op{\chi}(x)\op{\chi}(y)\rangle$) are not too large compared to other relevant scales (the so-called 'mean-field' regime). If this assumption is violated, equation (\ref{semiclassical-scalar}) becomes intractable, and physically the system may be better described as a superposition of macroscopically different backgrounds with fluctuations around each. In the nonequilibrium case this can happen dynamically if initially small fluctuations can grow over time, causing a breakdown over time of the mean-field assumption. Exploring the physics here beyond the mean-field regime is the domain of the comparatively new field of \emph{non-equilibrium quantum field theory} (see, \egc, \cite{bergesqftreview,calzettahubook} for details).

\section{Low-energy quantum gravity}\label{background-gravity}

Formally speaking, we can define a quantum theory of gravity using path integrals, just as we defined a scalar field theory above. We want a theory of the gravitational metric $g$ and some matter fields which we write schematically as $\phi$ (read $\phi$ as a scalar field if you like, though I intend it as a more general placeholder), and we assume an action of the form
\be
S[g,\phi] = \frac{1}{8\pi G}S^V[g] + S^M[\phi,g].
\ee
Here $S^V[g]$, the \emph{vacuum action}, has form
\be
S^V[g] =\int \dr{x^4}(R(x)  + \Lambda_0 )\sqrt{-g(x)} + \mbox{higher -order terms in }g
\ee
--- that is, it is the sum of the standard Einstein-Hilbert action, a cosmological-constant term, and unspecified higher-order terms in $g$. $S^M$ is a matter action, depending on both matter and metric: for a real scalar field it would have form\footnote{Arguably a non-minimally-coupled term $R \phi \sqrt{-g}$ should also be included: it violates no symmetry, and will be reintroduced by renormalization if not added initially.}
\be
S^M[\phi,g]=\int\dr{x^4} \left(\frac{1}{2}g^{\mu \nu}(x) \partial_\mu \phi(x) \partial_\nu \phi(x) - V(\phi(x))\right)\sqrt{-g(x)}.
\ee
The field theory is then specified by the path integral
\be
\bk{h_f,\phi_f}{h_i,\phi_i}= \int_{(g,\phi)|_{\mc{S}_i}=(h_i,\phi_i)}^{(g,\phi)|_{\mc{S}_f}=(h_f,\phi_f)}\mc{D}g\mc{D} \e{i S[g,\phi]}
\ee
where the boundary conditions indicate that we are integrating over fields between some initial and final 3-surfaces $\mc{S}_i$, $\mc{S}_f$, and that the fields on $\mc{S}_i$ must equal $(h_i,\phi_i)$ and similarly for $\mc{S}_f$.

This simple formal expression hides a host of conceptual and technical complications, of course, among them:
\begin{enumerate}
\item The path integral again requires regularization to be well-defined, and that regularization becomes more complex technically and conceptually due to the non-renormalizability of the gravitational action (see section \ref{ccp} for further discussion of this point).
\item The diffeomorphism invariance of $S$ means that a naive attempt to calculate the path-integral will overcount and integrate over fields that are related by a (small) diffeomorphism. That is conceptually undesirable (small diffeomorphisms are normally interpreted as purely redescriptive\footnote{High-energy physics does not treat \emph{large} diffeomorphisms --- those that are non-vanishing at infinity --- as redescriptive, and there is no requirement to avoid integrating over fields related by large diffeomorphisms; see \citeN{belotelvis,wallace-isolated-2} for discussion of the conceptual issues here.}) and leads to a pathological theory. The formal path integral $\mc{D}g \mc{D}\phi$ must be defined so as to integrate over only one element of each gauge orbit, which in actual calculations normally requires Faddeev-Popov ghosts, BRST methods, or other such heavy machinery (see \cite[ch.15-17]{weinbergqft2} and references therein).
\item Strictly speaking, the action must be supplemented by a boundary term to get a well-defined theory, due to the presence of second derivatives of the metric in the Einstein-Hilbert action; this has no classical consequences but can lead to non-perturbative quantum effects (see \cite[pp.749-751]{hawking-centenary} or \cite{padmanabhan-boundary} for a discussion). 
\item On a more conceptual level, diffeomorphism invariance leads to the notorious 'problem of time' (see \cite{butterfieldishamtime,ricklesQGreview} for discussion and further references), which can make it difficult even to interpret the formalism (at least in the cosmological case; for applications of quantum gravity to isolated systems like stars or black holes, asymptotically flat boundary conditions simplify matters). This shows up in the path integral in the fact that ---unlike in the scalar case --- I have not explicitly specified initial and final times. In isolated-system applications, we do actually need to specify those conditions, since the asymptotic boundary conditions make them well-defined; in the cosmological case, things are more complicated.
\end{enumerate}

But notwithstanding these subtleties and open questions, the theory is sufficiently well-defined that many calculations can be done in it, and we will see that it offers a unified starting point from which can be derived the various approximations discussed previously. In particular, we can apply the background-field method of the previous section, to express the metric as the sum of its expectation value $g_0$ and a new field $h$ which tracks fluctuations around that expectation value:
\be
g(x)=g_0(x)+ \sqrt{8\pi G}h(x)
\ee
(the reason for the $\sqrt{8\pi G}$ factor will be apparent shortly). Expanding the action around $g_0$ in this way, we get
\be
S[g_0+\sqrt{8\pi G}h,\phi]\equiv \frac{1}{8\pi G} S^V[g_0] +\frac{1}{\sqrt{8\pi G}}\left.\vbv{S^V}{g^{\mu\nu}}\right|_{g_0}\cdot h^{\mu\nu} \nonumber  +  S^M[g_0,\phi] + \Sigma_{g_0}[h,\phi]
\ee
and can then obtain a semiclassical analogue of (\ref{semiclassical-scalar}) for $g_0$:
\be \label{semiclassical-gr}
0  = \frac{1}{8\pi G}\left.\vbv{S^V}{g(x)}\right|_{g_0(x)} + \left\langle \vbv{(S^M[g_0,\cdot]+\Sigma_{g_0})}{g_0(x)}\right\rangle
\ee
where the expectation value on the right hand side is calculated for a field theory with fields $h,\phi$, defined on a spacetime background $g_0$ and with action
\be
S'_{g_0}[h,\phi]= S^M[g_0,\phi] +\Sigma_{g_0}[h]+ (\mbox{linear term in }h)
\ee
where, as before, the linear term can mostly be neglected. Since the stress-energy tensor in classical general relativity is defined as the variational derivative of the action with respect to the metric, we can interpret the second term in (\ref{semiclassical-gr}) as the expectation value of the stress-energy tensor for the $h$ and $\phi$ fields. Since the first, classical, term is the sum of the Einstein tensor, the cosmological-constant term, and higher-derivative terms, we can recognize (\ref{semiclassical-gr}) as a form of the semiclassical Einstein field equation.

It is instructive to explicitly expand $S[g_0+h,\phi]$ in powers of $\sqrt{8\pi G}$.\footnote{The attentive reader will observe that $G$ is dimensionful, which calls into doubt the interpretation of any such expansion. A more careful treatment (which I assume tacitly) would expand in powers of $(\sqrt{8\pi G}/L)$, where $L$ is the typical scale of the physical processes being analyzed (e.g., the inverse momenta of the particles). Since in our units $\sqrt{G}$ is the Planck length, this will be extremely small for low-energy phenomena and defines a meaningful expansion. The failure of this process when $L$ approaches Planck scales points to the nonrenormalizability of the theory and its consequent failure at those scales, discussed further in section \ref{high-energies}.  } We obtain
\begin{eqnarray}
S[g_0+\sqrt{8\pi G}h,\phi] &=& \frac{1}{8\pi G}S^V[g_0]\nonumber \\
&+& \frac{1}{\sqrt{8\pi G}}\left.\vbv{S^V}{g^{\mu\nu}}\right|_{g_0}\cdot h^{\mu\nu} \nonumber \\
&+& \left(S^M[g_0,\phi] + S^{GRAV}_{g_0}[h]\right) \nonumber \\
&+& \sqrt{8\pi G}\left(T^M_{\mu\nu,g_0}[\phi]h^{\mu\nu} + \frac{1}{3}T^{GRAV}_{\mu\nu,g_0}[h]h^{\mu\nu} \right) \nonumber \\
&+& O(8\pi G).
\end{eqnarray}
The first term here is the classical vacuum action, evaluated for the background field. Passing over the second term for the moment, the third term is the sum of the classical action defined on the background $g_0$, and the (free) \emph{graviton action} on that background:
\begin{eqnarray}
S^{GRAV}_{g_0}[h] &\equiv& \frac{1}{2}\left.\frac{\delta^2 S^V}{\delta g_c^2}\right|_{g_0}(h,h)\nonumber \\ &=& \frac{1}{4}\int\dr{x^4}\sqrt{-g_0}\nabla_\alpha h_{\mu\nu}\nabla_\beta h_{\rho\sigma}
[g_0^{\alpha\beta}g_0^{\mu\nu}g_0^{\rho\sigma} +2 g_0^{\alpha\sigma}g_0^{\mu\rho}g_0^{\nu\sigma} - (\nu\leftrightarrow \rho)]
\nonumber \\
&+& \mbox{cosmological-constant term} \nonumber\\
&+& \mbox{higher terms.}
\end{eqnarray}
Physical fields in QFT are normalized so as to avoid scale factors in front of the kinetic term for those fields, so this term is the reason for the $\sqrt{8\pi G}$ term in the definition of $h$.

Returning to the second term, we can now interpret it as a source term for the creation of single gravitons. Since such a process would shift the expected value of $\op{h}$ away from zero, we can now see that  the effects of this term will be exactly cancelled by other terms, and indeed this can be easily shown. At the technical level, the term just serves to cancel so-called `tadpole diagrams' in the evaluation of the path integral; interpretatively, it can be seen as compensating for quantum corrections to $\langle h \rangle$.

The fourth term is the lowest-order coupling of gravitons to matter and to each other. In each case, the gravitons couple linearly to the stress-energy tensor of the respective excitations: $T^M$ is defined in the standard way as the variational derivative of the matter action with respect to the (background) metric, and $T^{GRAV}$ is likewise the variational derivative of the free-graviton action. (The latter term is actually nonlinear because $T^{GRAV}$ itself depends on $h$, which also accounts for the $1/3$ term in that term: the variational derivative of $T^{GRAV}[h]_{\mu\nu}h^{\mu\nu}$ is $3T^{GRAV}[h]$, not $T^{GRAV}[h]$.) Higher-order terms encode higher-order couplings.

If we now evaluate the effective action to lowest order, dropping all positive powers of $8\pi G$, the semiclassical equation for $g_0$ becomes
\be
0= \frac{1}{8\pi G}\left.\vbv{S^V}{g_c(x)}\right|_{g_0(x)}  + \left\langle (T^M+T^{GRAV} ) \right\rangle
\ee
where the expectation value is calculated for a QFT consisting of the matter field $\phi$ on the background $g_0$ along with a field of free gravitons on that same background. We can recognize this as semiclassical gravity: a theory of quantum matter evolving on a classical spacetime background, with the backreaction of the quantum matter on the background given by Einstein's equation with the expected value of the stress-energy tensor as source term. (The graviton term is often omitted from presentations of semiclassical gravity, but --- at least in high-energy physics --- it is generally understood to be present in a careful formulation of the theory, and one periodically sees it mentioned explicitly (see, \egc, (Hu~\emph{et al}~\citeyearNP{hu-roura-verdaguer}) or \cite[p.8]{wallgsl}). It is normally negligible; an important, though theoretical, exception is black hole evaporation, where the presence of gravitons means that black holes evaporate by Hawking radiation even in pure (vacuum) quantum gravity.)

If we go to higher order, we include matter-graviton interactions, and hence indirectly self-gravitational effects of fluctuations in the matter field. It is at this level (again, often disregarding gravitons) that we can derive the linearized accounts of fluctuations which we have seen in inflation, and which also suffice to describe more homely fluctuation phenomena like the Page-Geilker experiment. As those fluctuations grow, the background field approach (whilst remaining formally exact) may become more and more calculationally intractable, and less and less useful as a description of the physics. Eventually one might need to abandon it and apply the more general methods of non-equilibrium quantum field theory; in the context of quantum gravity, these methods are known as \emph{stochastic gravity}\footnote{The term `stochastic gravity' is actually used both for the constructive approximation to an underlying QFT that I sketch here, and to more explicitly phenomenological approaches that seek to generalise the semiclassical equation by adding noise terms without requiring any principled microtheory. The two approaches are used concurrently by the same authors and I don't myself see any real conflict between them (they mirror the way phenomenological physics is developed in other areas). However, see \cite{mattinglyunprincipled} for discussion of stochastic gravity specifically from the phenomenological perspective.} (see \cite{hu-verdaguer-review} for a review of this relatively new, and growing, subject). But even in doing so, we are simply applying a different and more powerful set of approximation techniques to the same quantum field theory.

To conclude this section: we can define quantum gravity non-perturbatively as a quantum field theory, and extract both semiclassical gravity and linearized quantum gravity as approximations to that theory valid in certain regimes, using (at least at the broad schematic level) essentially the same techniques as we use in non-gravitational QFT. The path-integral formulation of low-energy quantum gravity provides a framework that unifies the various apparently-disconnected fragments of quantum gravity we have discussed so far and puts them under a single, albeit non-fundamental, theory.

\section{The scope of low-energy quantum gravity}\label{LEQG-scope}

For a theory as complicated as a quantum field theory (like QED, for instance), it rarely makes sense to say that any given experiment confirms the theory as a whole.\footnote{See \citeN{Cartwright1983} for further development of this point (though she draws more sweeping conclusions about the evidence for theories than I would want to endorse).} Rather, what we test in applications of these theories is invariably some particular approximation, regime of application, or limiting case to the theory. Some parts of the evidence for QED consist of evidence for its nonrelativistic electrostatic approximation; others support the existence of the photon; still others provide evidence for coherent interference between those photons. Collectively, the evidence for these various applications of QED provide very strong support for the theory as a whole, while always permitting the logical possibility that some different theory could underpin the selfsame class of approximations and idealizations that we actually test.

As an inevitably-partial classification of these approximation schemes for QED, we can consider asking two questions of the regime we want to analyze:
\begin{enumerate}
\item Is it relativistic? In the nonrelativistic regime, radiation can be neglected, long-range interactions can be treated as instantaneous, and particle creation and annihilation does not occur. In the relativistic regime, all these phenomena must be allowed for to varying degrees.
\item How quantum-mechanically does it treat the electromagnetic field? We can distinguish (i) a semiclassical regime, where the field is treated as sourced by the average charge distribution; (ii) a perturbative regime, where we need to treat the electromagnetic field as a superposition of field values but where the different terms in the superposition may be regarded as small fluctuations around a fixed background (normally the zero field); (iii) a macroscopic-superposition regime,\footnote{I don't intend to require any particular stance on the quantum measurement problem here. These macroscopic superpositions, at least for the purposes of this paper, can be interpreted \emph{a la} Everett, but can also be thought of as probability distributions over actually-occuring macroscopic outcomes, where those probability distributions are interpreted via some unspecified hidden-variable theory or dynamical collapse mechanism or as part of a broader Copenhagen-style reading of quantum mechanics.} where we need to allow the quantum state to evolve into a superposition of macroscopically distinct backgrounds. In case (ii) we can further distinguish (iia) a perturbative/incoherent regime, where interference between terms in the superposition can be neglected and the state behaves like a classical mixture; (iib) a perturbative/coherent regime, where interference cannot be so neglected. (In principle we could make the same distinction for (iii); in practice, decoherence normally means that we cannot expect to observe interference between macroscopically distinct terms in a superposition.)
\end{enumerate}
One way to see the strength of the evidence for QED is to note that there is extensive empirical support for all eight regimes derived from the various combinations of answers to these questions. (I don't intend the following list to be remotely exhaustive.) The semiclassical regime for QED leads to the Hartree-Fock approximation, very extensively applied to relativistic and nonrelativistic atoms and molecules. The perturbative/incoherent regime includes Coulombic scattering theory at nonrelativistic energies, and the standard photon/electron scattering theory of QED at relativistic energies. The perturbative/coherent regime includes multi-atom and multi-ion entanglement in nonrelativistic physics, and the classic photon two-slit experiment and its many generalizations in relativistic physics. And spontaneous breaking of symmetry , and its spontaneous onset as systems are cooled below their critical temperature (in, \egc, superconductors or Bose-Einstein condensates) requires the macroscopic-superposition regime, including (when we want to consider electromagnetic radiation incident on a superconductor, for instance) relativistic applications of that regime. (See Table \ref{table-qed} for a summary.)

\begin{table}
\caption{Evidence for QED in various regimes}
\label{table-qed}
\begin{center}
\begin{tabularx}{0.8\textwidth} { 
  | >{\raggedright\arraybackslash}X 
  || >{\raggedright\arraybackslash}X 
  | >{\raggedright\arraybackslash}X | }
 \hline
 Quantum superpositions & Non-relativistic & Relativistic \\
\hline\hline
Semiclassical & Hartree-Fock methods (nonrelativistic atoms) & Hartree-Fock methods (relativistic atoms) \\
\hline
Perturbative / Incoherent & non-relativistic (Coulomb) scattering & electron-photon scattering \\
\hline
Perturbative / Coherent & multi-atom and multi-ion entangled states & photon two-slit experiment \\
\hline
Macroscopic & symmetry breaking on cooling: Bose-Einstein condensates & symmetry breaking on cooling: superconductors \\
\hline
\end{tabularx}
\end{center}
\end{table}
Performing the same exercise for low-energy quantum gravity  (table \ref{table-qg}) shows, unsurprisingly, that it is less well confirmed empirically than QED.  (See Table \ref{table-qg} for a summary.)  We have really abundant evidence from astrophysics and cosmology  for the semiclassical regime both at nonrelativistic  (section ~\ref{examples1}) and relativistic (section ~\ref{examples2}) energies, pretty good evidence (section ~\ref{nr-fluctuations}) for the nonrelativistic part of the perturbative/incoherent regime from the Page-Geilker experiment and more general considerations of Solar-system-scale physics, and some evidence (section ~\ref{inflaton}) for the relativistic part of that regime from primordial cosmology. Stochastic gravity methods allow us to probe the macroscopic regime, mostly in the context of early-universe structure formation (and the boundary between those methods and perturbative treatments of fluctuation formation is not completely sharp) but to my knowledge have not yet produced really robust predictions. (This regime only really makes sense relativistically: in the domain of nonrelativistic gravity, perturbative methods seem likely to capture the full content of the theory.)

\begin{table}
\caption{In which regimes is low-energy quantum gravity observationally supported?}
\label{table-qg}
\begin{center}
\begin{tabularx}{0.8\textwidth} { 
  | >{\raggedright\arraybackslash}X 
  || >{\raggedright\arraybackslash}X 
  | >{\raggedright\arraybackslash}X | }
 \hline
 Quantum superpositions & Non-relativistic & Relativistic \\
\hline\hline
Semiclassical & \textbf{strongly supported} (planets, white dwarfs) &  \textbf{strongly supported} (stars, neutron stars, supernovae, early-universe cosmology) \\
\hline
Perturbative / Incoherent & \textbf{strongly supported} (Page-Geilker, general considerations of solar-system physics) &  \textbf{some support} (quantum-fluctuation models of the CMB) \\
\hline
Perturbative / Coherent & \textbf{limited support so far} but experiments are ongoing & \textbf{no support} and little prospect of direct testing \\
\hline
Macroscopic & \textbf{not applicable} (this regime can be analyzed perturbatively without loss) & \textbf{developing} (stochastic-gravity methods probe this regime but no decisive confirmed predictions) \\
\hline
\end{tabularx}
\end{center}
\end{table}

At present, we have little evidence supporting applications of LEQG in the perturbative/coherent regime, which is significant because it is this regime which distinguishes unitary quantum gravity from variants in which superpositions of mass distributions cause wavefunction collapse, as has been advocated by D{\'i}osi~\citeyear{diosi1987,diosi2014},  Penrose~\citeyear{penrose1969,penrose-1996,penrose-2014}, Stamp~\citeyear{stampqg} and others. For that very reason there has been considerable interest in probing that regime, at least nonrelativistically, by attempting to place a relatively-massive object in a superposition and to test the coherence of that superposition via interference experiments. Those experiments are challenging, but not impossible with current technology: recent experimental work by Donadi~\emph{et al}~\citeyear{donadi2021} appears to rule out at least the simplest form of the Di{\'o}si-Penrose model of gravitational collapse, and in doing so to corrobate at least a part of the perturbative/coherent nonrelativistic regime. We can optimistically hope for LEQG to be more systematically confirmed --- or falsified --- in that regime in the comparatively near future, with recent proposals for experiments on self-gravitating entangled systems~\cite{marlettovedralgravity,christodoulourovellidetectinggravity}. Probing more strongly relativistic aspects of the perturbative/incoherent regime looks far more challenging (performing the two-slit experiment with gravitons, for instance, is ridiculously beyond any current or foreseeable gravity-wave-detection technology) so that it seems likely any tests of this regime will have to be indirect, probably via cosmology.And no breakdown of this kind can hope to be exhaustive: no doubt the theory will sooner or later be put to the test in regimes that defeat my categorization.

In summary: we know how to apply LEQG in a fairly wide, and growing, range of physical situations. In some of those situations there is extremely good evidence for its accuracy; in others there is some evidence; in others there is the prospect of getting evidence in the near future; in others still, there is little prospect. Certainly it would be premature to treat it as confirmed to anything like the degree of QED; certainly, it is well worth striving to confirm it further observationally or experimentally, in novel regimes. And the very concreteness of the predictions that the theory makes serve to sharpen the task of any such observation or experiment. Low-energy quantum gravity shows us that it is too pessimistic to suppose that quantum gravity must remain a theoretical endeavour: if the predictions of LEQG are often hard to test, still they are made at energy scales far more accessible than the Planck scale. Still, the theory provides a reasonably satisfactory unifying basis for all those quantum-gravitational applications we currently have and in doing so is reasonably well confirmed by experiment. And --- perhaps with the exception of the various gravitational-induced collapse models --- it is difficult readily to see what alternative could provide a unified underpinning for those parts of LEQG which are strongly supported by evidence.

For all that, it is not fundamental. We have strong reasons to believe that the theory will fail at sufficiently high energies, and that it has other issues --- notably the infamous Cosmological Constant Problem --- that may point to problems over and above that failure. I discuss the latter in section \ref{ccp}; before that, I discuss the limitations on LEQG that arise from its nature as a nonrenormalizable effective field theory.

\section{Quantum gravity at \emph{high} energies}\label{high-energies}

Any quantum field theory is defined with an (implicit or explicit) high-energy cutoff. At the formal level, the cutoff serves to make the path integral (tolerably) well-defined, and to regulate the divergences in perturbative calculations. At the physical level, the cutoff represents a division between those degrees of freedom that the QFT explicitly models and those which it does not. There is in general no requirement that the cutoff represents a physical scale at which the physics fundamentally changes: given a QFT with a cutoff at some energy $E$, we can always integrate out degrees of freedom between $\lambda E$ and $E$ and obtain a new QFT with a cutoff at energy $\lambda E$. But if we ask the inverse question --- how \emph{high} can the cutoff be set before the QFT breaks down? --- then we get different answers for different QFTs.

The best behaved cases are \emph{asymptotically safe}  QFTs, of which the physically most significant examples are pure non-Abelian gauge theories like QCD. Here the cutoff can be set arbitrarily high, and the interaction strengths in the theory become lower as particle energies increase, tending to zero or to some finite value. For these QFTs it is at least a live possibility to suppose that there is a well-defined continuum limit of the theory which could be defined in a mathematically rigorous way (though the physical significance of doing so can be questioned; cf \cite{wallacecritique}). 

Next there are renormalizable but non-asymptotically-free theories; here some interactions actually increase in strength at higher energies, and there is good reason to suspect that \emph{very} high values of the cutoff eventually cause the interaction strengths to diverge and the theory to become ill-defined. However, the energy scales on which these `Landau poles' occur are higher by far than any physcially relevant energy scales (higher by far than the Planck scale, for instance). QED and (the $U(1)$ and Higgs sectors of) the Standard Model are of this form: we have reason to doubt that they have a genuine continuum limit, but they can be well defined at energy levels far beyond those for which external reasons tell us to expect them to lose empirical validity.

The least well behaved cases are \emph{nonrenormalizable} theories. Here the theory imposes a much lower and sharper bound on the value of the cutoff. In a nonrenormalizable theory, if we try do to calculations that approach that bound, we will first lose predictive power (each new order of perturbative calculation will involve new constants, knowable only from experiment, whose effects are not sharply suppressed unless we calculate at energies far below the bound). If we persist, unitarity itself will break down. For theories of this kind, the bound on the cutoff represents a directly physical, internal-to-the-theory constraint on the energies at which it is applicable. Various low-energy sectors of the Standard Model --- notably weak-interaction-induced lepton scattering and pion theory, as well as the electroweak sector as a whole without the Higgs boson --- are well described by nonrenormalizable QFTs; more importantly for our purposes, low-energy quantum gravity is non-renormalizable.

In the days before the effective-field-theory paradigm, the cutoff was widely regarded as a purely formal device, ideally to be set to infinity at the end of a calculation. From this perspective,\footnote{It is probably fair to say that `this perspective' is a little over-simplified as a description of the actual history: physicists in the pre-effective-field-theory era had more subtle and nuanced views of the high-energy cutoff than modern physics textbooks might suggest. See \cite{li-qed} for further discussion.} the only really satisfactory QFTs are the aymptotically-free ones. Renormalizable theories with Landau poles can be tolerated but induce unease. And non-renormalizable theories are pathological and must be disregarded.

But if a quantum theory is thought of as an effective field theory, nonrenormalizable theories lose their stigma; instead, they convey valuable information about the energy levels at which new physics can be expected. The four-fermion theory of weak lepton scattering must be cut off at energies of $\sim 100 \mathrm{GeV}$, pointing to new physics at around that scale; that `new physics' turns out to be the $W$ and $Z$ bosons which mediate the weak interaction, with masses of $80 \mathrm{GeV}$ and 92 $\mathrm{GeV}$ respectively. (And the creation of these particles in high-energy scattering of leptons provides a directly physical explanation of unitarity violation in the lepton sector). The Higgs-free part of the Standard Model must be cut off at energies of a few hundred GeV to be well-behaved; hence, whether or not the Higgs boson was found in the LHC, physicists could be confident that \emph{some} new physics would be found.\footnote{For a review of the physics here, see \cite[chs.19,21]{weinbergqft2} or (Donoghue~\emph{et al}~\citeyearNP{dynamicsofstandardmodel}).} And LEQG must be cut off at the Planck length (and is calculationally reliable only for energies well below the Planck length), which tells us that new physics will be needed to describe systems at Planckian energy scales.\footnote{This is the standard view in physics on renormalizability of LEQG; there is, however, a longstanding research program (`asymptotic safety') to explore whether non-perturbative methods might after all make the theory definable at all scales. (The original idea is due to \citeN{weinbergasymptotic}; see \cite{niedermaier,reutersaueressig} for reviews). For my purposes this caveat doesn't matter much: as I explain below, even if there is an extension of LEQG in this way, its interpretation at Planckian scales will deviate wildly from the relatively-tame QFT interpretation of LEQG.}

There is also a more directly physical argument why LEQG, and the methods of QFT in general, ought to fail at around the Planck length. (Here I follow \citeN[pp.95-6]{susskindlindesay}.) In QFT, the effective minimum size $r(m)$ of an quantum excitation of mass $m$ is set by its Compton wavelength --- $r(m)\geq 1/m$, in the units we are using. But in general relativity, there is another constraint on the minimum size of a particle: it must lie outside its own Schwarzschild radius, or else an event horizon will form around it and (at least as regards external physics) it will behave as a black hole of that radius. This constraint imposes $r(m) \geq 2Gm \sim (l_P/m_P) m$, where $l_P$ and $m_P$ are the Planck length and mass, respectively (note that as we should expect from this classical calculation, their ratio does not depend on $\hbar$, ).

These constraints cross over at the Planck scale: a particle with a mass greater than $\sim m_P$ has a Compton wavelength contained within its own Schwarzschild radius. The Planck length is thus a cutoff on the energies of elementary (\iec, non-compound) excitations of quantum fields. If a spacetime description is applicable at all at scales below the Planck length, it cannot be a description in terms of excitations of quantum fields, of the kind we are used to from QFT. 

We can see this in another way by considering scattering theory. In QFT, the energies of the products of a collision with center-of-mass energy $m$ also have energy $\sim m$. But if we collide elementary particles with a center-of-mass energy $>m_P$, they will form a black hole, which will decay by emitting Hawking quanta of energy $\sim m_P/m$. In other words, above the Planck length the energy of collision products actually begins to decrease again. As Susskind and Lindesay put it (pp.95-6): 
\begin{quote}
Thus we see that a giant ``Super Plancktron Collider'' (SPC) would fail to discover fundamental length scales smaller than the Planck scale $l_P$, no matter how high the energy. In fact, as the energy increased we would be probing still larger scales\ldots
\end{quote}
What can we say about quantum gravity at Planckian scales, given this failure of LEQG at those scales? Let's distinguish two sorts of theory:
\begin{enumerate}
\item An \emph{ultraviolet quantum gravity} theory (UVQG) is any theory from which LEQG can be derived as an effective field theory valid in some limited regime of the UVQG.
\item A \emph{correct quantum gravity} theory is any theory which actually gets the physics right in some regime which includes (i) the domain of empirical validation of LEQG and (ii) at least some part of the Planckian regime.
\end{enumerate}
The two are not logically equivalent. Empirical adequacy for a quantum theory of gravity, at present, only requires it to reproduce the predictions of LEQG in certain domains: the semiclassical domain, the perturbative/incoherent domain; soon, perhaps, the nonrelativistic part of the perturbative/coherent domain. (One sometimes hears it said that the theory need only reproduce the predictions of non-gravitational QFT and classical GR; sections \ref{examples1}--\ref{inflaton} should make clear why this is far too weak a demand to make.) Most physicists working on quantum gravity set themselves the more demanding task of reproducing most or all of LEQG, notably including its predictions of black hole evaporation, which are unsupported by any observational evidence to date: that is, they are searching for a UVQG theory. In doing so, they go beyond what evidence requires of them. That said, their strategy seems well-motivated pending a low-energy alternative to LEQG. The problem of reproducing those fragments of LEQG that are directly supported by evidence is just too open to provide much guidance, whereas reproducing all of LEQG is a very strong constraint on theorizing.

Conversely, there is no obvious reason why a UVQG need be a (let alone the) \emph{correct} quantum gravity theory. If we had two inequivalent ultraviolet completions of LEQG, nothing in LEQG would allow us to distinguish them. And since we have virtually no observational access to regimes in which LEQG is not valid, prospects would be dim for any means of distinguishing the two. However concerning this might be, however, it is a rather theoretical problem at the current state of development of quantum gravity, where our problem is not that we have two or more UVQGs but that we do not have even one that is generally accepted to be successful in reproducing LEQG in all the domains in which it can be applied.

As a consequence, and notwithstanding these constraints, when physicists speak of a `quantum theory of gravity' they normally mean an ultraviolet completion of LEQG (when they do not mean LEQG itself). And the basic requirement that the theory  \emph{is} an ultraviolet completion of LEQG does tell us at least a little about it.

Firstly, it is not a QFT, or at any rate not a QFT interpretable in anything like the way we normally interpret QFTs (including LEQG). The `excitations on a background spacetime' picture of QFTs, as we have seen, breaks down at Planckian scales; well below those scales, LEQG works fine and there is no obvious extra work for a UVQG to do. If QFT methods remain applicable for Planck-scale physics, they must be applicable in a radically different way. (The AdS/CFT correspondence, arguably our best candidate so far for UVQG, illustrates how this might happen: quantum-gravity phenomena in the interior of asymptotically-AdS spacetime are describable indirectly via a QFT on the conformal boundary of that spacetime.)

Secondly, by the same token a UVQG will be ultraviolet \emph{complete}, and not just another effective field theory. The reason is not that we hubristically expect our theory to apply at arbitrarily high energy scales, but that the idea of an energy scale in the normal way we use it loses meaning beyond the Planck scale. To say that a QFT is ultraviolet-complete, mathematically, is to say (among other things) that the Fourier-transformed four-point function remains well-behaved at arbitrarily large values of input wavenumber. But the normal physical interpretation of that four-point function as encoding particle scattering cross-sections at successively higher energies becomes meaningless at Planckian scales. If our UVQG is restricted in its domain of application, that restriction will need to be statable in novel ways and not by our usual considerations of energy scale.\footnote{Though see \cite{crowtherlinnemann} for a dissenting view.} (Again, AdS/CFT correspondence offers an illustration of how this could occur: the `asymptotically AdS' boundary conditions are normally interpreted as a way of putting a system in a box, so that AdS/CFT gives a quantum-gravity description of isolated systems but is inadequate for cosmology.)

Thirdly, it will need to be a `theory of everything' in the sense that it will need to incorporate the non-gravitational degrees of freedom described by the Standard Model, not just purely-gravitational degrees of freedom. The reason is comparatively mundane: there is just no clean separation of these degrees of freedom at Planckian scales. The annihilation of two gravitons to produce leptons and antileptons, for instance, is a risibly unimportant process in LEQG, but becomes significant as we approach the Planck scale. There is no reason to expect a quantum theory that reproduces just the vacuum sector of LEQG to be empirically adequate. This is, however, neutral as to whether a UVQG includes a more thoroughgoing unification of fundamental forces (as string theory attempts to do) or more conservatively maintains the gauge structure of the Standard Model (as loop quantum gravity does).

Finally, and most importantly, it will be highly underconstrained by observational evidence except insofar as that evidence is evidence for LEQG. This is what is meant when experts talk about the near-absence of evidence for quantum gravity: it is not that they are ignorant of LEQG but that they mean ultraviolet quantum gravity, and are speaking of evidence that goes beyond evidence for LEQG. And that is, indeed, difficult in the extreme to obtain, requiring energy levels which seem to occur in our Universe only behind the screen of the plasma of the very-early Universe or the still more opaque screen of a black hole's event horizon. From this point of view, the success of LEQG stands to the search for UVQG much as the success of the Standard model stands to the search for post-Standard-Model physics: we have excellent grounds to think that the theory is not the last word, but nonetheless it is so successful as to make it difficult to go beyond it.

So it is exciting as well as puzzling that LEQG appears to contain a severe anomaly that might point towards new physics: the mysteriously low value of the observed cosmological constant.

\section{The cosmological constant problem}\label{ccp}

Suppose that we want to apply low energy quantum gravity in the background-field regime and for some comparatively tame region of the background spacetime. Specifically, let's suppose that the scale of spacetime curvature in this region is $\gg 10^{-15}\mathrm{m}$, that the matter field can be described in this region as a small perturbation away from some thermal-equilibrium state, and that that state is either the vacuum or at any rate is at a temperature $\ll 10^{14}\mathrm{K}$. (These are not unduly demanding constraints. The curvature scale near the event horizon of an astrophysical black hole is $\sim 10^{4}\mathrm{m}$; the temperature in the core of a supergiant star just before becoming a supernova is $\sim 10^{12} \mathrm{K}$.) A particle of energy $>100 \mathrm{MeV}$ in this region will effectively behave as if in the Minkowski vacuum (the Compton wavelength of a 100-MeV particle is $\sim 10^{-15} \mathrm{m}$; the typical energy of a particle in an equilibrium state at temperature $10^{14} \mathrm{K}$ is $\sim 100 \mathrm{MeV}$). Or put another way, at least for excitations on scales of $\sim 100 \mathrm{MeV}$ or higher, we should be able to make local calculations as if in the Minkowski vacuum.\footnote{Why not the de Sitter vacuum, given the positive observed cosmological constant? You can use that if you like, but it doesn't matter (and is mildly more annoying technically). The observed value of the cosmological constant corresponds to a curvature scale of $\sim 10^{26}\mathrm{m}$, or $10^{38} \times $ the Compton wavelength of our 100-MeV particle. The local physics just doesn't care about the difference between this level of curvature and exact flatness, not unless we're looking at cosmological-wavelength excitations.}

Let's now try to apply the rules of the background-field method: for self-consistency (and ignoring higher-order terms in the gravitational action), we need
\be
\frac{1}{8\pi G}(G_{\mu\nu}+ g_{\mu\nu} \Lambda_0) = \left\langle \op{T}_{\mu\nu}\right\rangle
\ee
where the expectation value is calculated for the quantum state of the matter fields (and, in principle, the gravitons) on the background. What we would hope to find is that there are two contributions to that stress-energy expectation value: one due to the thermal energy of the finite-temperature local equilibrium state (if we are indeed at finite temperature) and one due to quantum excitations. The latter term should be treatable as a sum of particle energies if the excitations are on lengthscales short compared to the curvature.

We do indeed find that, but we also find something much less welcome. For the moment, let's suppose that the calculation is carried out on a Minkowski background. We then get a series of very large terms proportional to the Minkowski metric. By far the largest of these terms, is of order $(E_{\mbox{cutoff}})^{4}$, where $E_{\mbox{cutoff}}$ is the cutoff energy scale we are using to regularize LEQG (normally assumed to be around the Planck scale, $\sim 10^{19} \mathrm{GeV}$); this term can be understood as the zero-point energy density of the vacuum. The second term is of order $(E_{EW})^4$, where $E_{EW}\sim 100 \mathrm{GeV}$ is the scale of electroweak symmetry breaking; it can be understood as the energy density of the Higgs condensate, due to the spontaneous breaking of electroweak symmetry. The third term is of order $(E_{CH})^4$ where $E_{CH} \sim 1 \mathrm{GeV}$ is the scale of approximate chiral symmetry breaking in the QCD sector of the Standard Model. (Our stricture that $T\ll 10^{14}\mathrm{K}$ ensures that these broken symmetries remain broken for our quantum state.)

(As a calculational matter, the two condensate terms occur at tree order in the calculation, and the zero-point term occurs at one-loop order. There will be one-loop quantum corrections to the condensate terms, and higher-order corrections to all the terms, but they do not change the basic magnitudes of the terms. They do somewhat blur the clean distinction between the three contributions, but not in a way that really matters for the argument here.)

The zero-point term arises due to fluctuations on scales far smaller than the assumed curvature of our local region, and the other two terms are due to potential effects insensitive to curvature, so we can take our Minkowskian calculation over to that region just by replacing the Minkowski metric with the local background metric. The semiclassical equation can then be written as
\be\label{semiclassical-cc}
\frac{1}{8\pi G}(G_{\mu\nu}+ g_{\mu\nu} \Lambda_0) = \left\langle \op{T}^{\mbox{IR}}_{\mu\nu}\right\rangle +\frac{1}{8\pi G}g_{\mu\nu} (\Lambda_{ZP} + \Lambda_{EW} + \Lambda_{CH}) 
\ee
or (rearranging)
\be\label{semiclassical-cc2}
\frac{1}{8\pi G}\left( G_{\mu\nu}+ g_{\mu\nu} (\Lambda_0 - \Lambda_{ZP} - \Lambda_{EW} - \Lambda_{CH})\right) = \left\langle \op{T}^{\mbox{IR}}_{\mu\nu}\right\rangle
\ee
where $\Lambda_{ZP}/8\pi G$, $\Lambda_{EW}/8\pi G$, and $\Lambda_{CH}/8\pi G$ are respectively the zero-point, Higgs condensate, and chiral condensate contributions to the stress-energy density, and where (translating into SI units for clarity),
\begin{eqnarray}
\Lambda_{ZP} & \sim & 10^{97} \mathrm{kg}\,\mathrm{m}^{-3} \nonumber \\
\Lambda_{EW} & \sim & 10^{29} \mathrm{kg}\,\mathrm{m}^{-3} \nonumber \\
\Lambda_{CH} & \sim & 10^{21}  \mathrm{kg}\,\mathrm{m}^{-3}.
\end{eqnarray}
The first term on the right-hand-side of equation (\ref{semiclassical-cc}), the `infrared' part of the stress-energy tensor, encodes contributions from the finite-temperature background, from particle excitations on scales small compared to the curvature, and potentially from other excitations on longer scales to which the particle concept cannot be applied; however, on basic dimensional-analysis grounds it will be far smaller than any of the other three terms.

 Now, the observationally-supported version of the semiclassical equation is
\be\label{semiclassical-observed}
\frac{1}{8\pi G}(G_{\mu\nu}+ g_{\mu\nu} \Lambda_{obs}) = \left\langle \op{T}^{\mbox{IR}}_{\mu\nu}\right\rangle 
\ee
where $\Lambda_{obs}$ is the observed cosmological constant. Comparing equations (\ref{semiclassical-cc2}) and (\ref{semiclassical-observed}), we get
\be \label{lambda-sum}
\Lambda_{obs}=\Lambda_0 - (\Lambda_{ZP} + \Lambda_{EW} + \Lambda_{CH}).
\ee

In one sense, there is no difficulty solving this equation. $\Lambda_0$ --- the \emph{bare} cosmological constant, in particle-physics terminology --- influences observable quantities only through equation (\ref{lambda-sum}), and so we can take (\ref{lambda-sum}) simply as fixing the bare cosmological constant as a function of the observable parameters and the high-energy cutoff, as usual in QFT.

On the other hand, 
\be
\Lambda_{obs}\sim 10^{-26} \mathrm{kg}\,\mathrm{m}^{-3},
\ee
which is about 120 orders of magnitude smaller than the zero-point energy density and about 50 orders of magnitude smaller even than the chiral-condensate energy density. So the small value of the observed cosmological constant we observe today requires the bare cosmological constant --- or, equivalently, the bare parameters of the theory as a whole --- to be fine-tuned to an extraordinarily precise value, to one part in $10^{120}$ or more.\footnote{The relationships between bare and observed parameters in particle physics are \emph{not} generally fine-tuned in this way; the exception is the Higgs mass, where fine-tuning concerns constitute the closely-related \emph{hierarchy problem} of the Standard model.} This is the cosmological constant problem (CCP).\footnote{The literature on the CCP is too vast to systematically reference here. The classic statement of the problem remains \cite{weinbergcosmologicalconstant}; more recent introductions/reviews include \cite{carrollccp,polchinskiccp,burgessccp,padillaccp}; \cite{rughzinkernagel} is a good non-technical introduction. }

Whether fine-tuning to this degree is a problem at all, and if so how it might be resolved, lies beyond the scope of this article. Maybe we should be relaxed about it, since after all $\Lambda_0$ is unobservable; maybe we can solve it with subtle short-distance/long-distance effects which correct the effective-field-theory framework; maybe a resolution needs to draw on anthropic arguments or something still more radical.\footnote{For scepticism about the significance of the fine-tuning see \cite{bianchirovelliconstant,hossenfeldernaturalness}; 
for anthropic considerations see \cite{polchinskiccp}; see also \cite{wallacenaturalness} and references therein.} But the fine-tuning arises as a direct result of comparatively simple calculations within LEQG.

Recent critical discussions of the CCP (\egc, \citeNP{saunderscosmologicalconstant}; Koberinski~\citeyearNP{koberinskicosmologicalconstant,koberinskivacuum}; \citeNP{schneiderccp})
have generally presented the problem quite differently, as relying on multiple conceptual assumptions and speculations about as-yet-undeveloped physics. On this approach, we first have to establish that the zero-point energy is real at all, perhaps by consideration of the Casimir effect; then we have to argue that it gravitates; then we have to argue that the way it gravitates is via the semiclassical equation. At each step, it is suggested, the problem turns on some specific conceptual choices, and by the end of the story one is left with the impression that the existence of the CCP is speculative at best. But if the CCP is thought of as a problem inside LEQG, no conceptual steps are required. All one has to do is derive the semiclassical equation via the background-field method and then calculate the terms in it.

(That calculation can itself be challenged. \citeN{koberinskivacuum} argues that the calculated value of the zero-point energy, depending sensitively on the cutoff as it does, cannot be treated as a physical prediction of the theory; 
Koberinski and Smeenk (in forthcoming work) also raise concerns 
about the applicability of the `Minkowskian approximation' which I appealed to above when applied to even gently curved spacetimes. I find both objections unpersuasive. As to the first: the fact that the zero-point energy is extremely large is \emph{not} sensitive to the cutoff; nor is the fact that the bare cosmological constant requires extreme fine-tuning. And we cannot simply drop a term in a calculation because we find it unreliable, not least because we need to extract a finite, physically-significant temperature-dependent component from the zero-point term. As to the second: while it is certainly worth being concerned about various ways in which QFT calculations might subtly fail, the Minkowskian method is used pretty widely in calculating equations of state in astrophysics and cosmology, and more generally, the basic idea of the separation-of-scales approach which lets us treat short-wavelength phenomena as insensitive to long-distance features of the background is totally ubiquitous in physics. But in any case, we can sidestep these concerns entirely just by appealing to the Higgs-condensate and chiral-condensate contributions to the stress-energy density, which are insensitive to the high-energy cutoff and can be calculated at tree level without needing to appeal to any Minkowskian approximation to the field modes.) 

If the CCP is a consequence of low energy quantum gravity, doesn't it give us reason to worry that something is wrong with that theory? And couldn't we avoid the CCP just by abandoning LEQG? Yes, and yes. The latter is trivial: \emph{of course} we can solve a deep conceptual problem in our best current theory, whatever it is, by discarding it. (Throw away quantum theory and \emph{voila}, no quantum measurement problem!) The former is pretty much universally recognized in the physics community: Weinberg's famous paper on the cosmological constant problem \cite{weinbergcosmologicalconstant} begins ``physics thrives on crisis'' and analogizes the CCP to other historical crises that have led to new physics. But a successful resolution of the CCP requires not just the bare observation that a theory other than LEQG might not have a CCP, but an actual construction of such a theory so as to preserve the observational successes of LEQG. And the challenge is all the more stark because the problem arises already in the semiclassical regime of LEQG, which is exactly the regime where observational evidence is strongest. 

We can state the CCP as a simple challenge. Suppose that you are tasked with studying astrophysics or cosmology in a regime where a quantum-field-theoretic description of matter is needed (but where fluctuations in the matter distribution are not significant): perhaps the regime is the interior of neutron stars; perhaps the core of a supernova; perhaps the quark/gluon plasma of the reasonably-early Universe. To make any progress you will need to calculate an equation of state for the matter. If you calculate the equation of state using the methods of quantum field theory to derive the expected stress-energy tensor, you will immediately encounter very large contributions proportional to the metric from the Higgs and chiral condensates, and will encounter even larger contributions once you proceed beyond leading order, and you will need to cancel them out using a very precisely tuned bare cosmological constant in order to match observed results; in other words, you will be confronted with the CCP. If you plan to perform your calculation without using QFT in this way, then you had better be able to say what your plan is.

\citeN{kuhn} famously argued that scientific theories are replaced not because clear experiments falsify them, but through the slow build-up of `anomalies' in our extant theories --- theoretical or empirical problems in the theory. Confronted with an anomaly, practicing scientists do not simply abandon an otherwise-successful theory; they note the anomaly and work around it. Sometimes anomalies are resolved through deeper understanding of the theory, but at other times, they point the way to an eventual replacement of the theory. The CCP is a Kuhnian anomaly in  low-energy quantum gravity: absent some better theory, it is no reason to abandon LEQG, but it points to deficiencies either in LEQG or in our understanding of it, and may help us obtain that better theory.

\section{Conclusion}\label{conclusion}

If `quantum gravity' means `a physical theory which combines general relativity and quantum theory', or `a physical theory which has classical general relativity and non-gravitational QFT as limiting cases or partial approximations' then we have such a theory, and have had it for decades. It is the straightforward, conservative theory one obtains by treating GR as an effective field theory, quantized through path-integral means. There is much we do not understand about even this theory, but we understand enough to know that it is applicable to a wide variety of observational contexts, many of which are accessible with current observational and experimental methods. Indeed, the theory has been widely applied in many such contexts, and those applications match the observational data; those applications are invariably through one or more approximation schemes, but that is true of virtually any application of a complicated physical theory to concrete examples. Much (most?) of contemporary astrophysics and cosmology relies on our ability to synthesise quantum-mechanical and gravitational phenomena, and low-energy quantum gravity provides such a synthesis which seems fairly satisfactory in its own terms; its nonrenormalizability, and its inevitable failure at Planckian scales, is unexceptional in the era of effective field theories. And since it includes as limiting cases both (semi)classical and phenomenological general relativity, and the Standard Model of particle physics, it has a reasonable claim to be the most widely applicable physics theory ever obtained, with a scope covering almost every experiment and observation we could hope to make. But by the same token, the theory makes many predictions which have yet to be tested but which lie well below the nigh-inaccessible Planck scale. 

If `quantum gravity' means instead `an ultraviolet completion of low-energy quantum gravity', or `a theory of gravity that applies at Planckian scales', then it is true that we do not have such a theory, and also true that very little empirical evidence bears on such a theory --- except insofar as it also bears on low-energy quantum gravity. As such, reproducing low-energy quantum gravity --- and not simply non-gravitational QFT and classical general relativity --- is generally taken as the success condition for any such theory. That does not, and should not, rule out attempts to construct a Planck-scale quantum gravity theory which differs from low-energy quantum gravity in a yet-unexplored regime. But any such attempts still have large fragments of low-energy quantum gravity, going well beyond non-gravitational QFT and classical general relativity --- that they need to reproduce to match observations.

The Cosmological Constant Problem arises within low-energy quantum gravity via straightforward calculation: the parameters in the theory need to be tuned to a precision of one part in $10^{120}$ to reproduce the observed cosmological constant. This does not falsify low-energy quantum gravity, but it is hard to explain that fine-tuning within the established paradigm of effective field theory. The cosmological constant problem is thus one of the few clues we have as to what might lie beyond this extraordinarily powerful and successful, yet ultimately limited, theory.

\section*{Acknowledgements}

Thanks to Adam Koberinski, Simon Saunders, Mike Schneider, and audiences in Oxford and Pittsburgh for useful discussion; thanks also to Adam for comments on a previous draft.


\begin{thebibliography}{}

\bibitem[\protect\citeauthoryear{Ade et~al.}{Ade et~al.}{2018}]{bicepreview}
Ade, P. et~al. (2018).
\newblock Constraints on primordial gravitational waves using {P}lanck, {WMAP}
  and new {BICEP2}/{K}eck observations through the 2015 season.
\newblock {\em Physical Review Letters\/}~{\em 121}, 221301.

\bibitem[\protect\citeauthoryear{Arnett}{Arnett}{1996}]{arnett}
Arnett, W.~D. (1996).
\newblock {\em Supernovae and nucleosynthesis: an investigation of the history
  of matter, from the Big Bang to the present}.
\newblock Princeton: Princeton University Press.

\bibitem[\protect\citeauthoryear{Ashby}{Ashby}{2003}]{ashbyGPS}
Ashby, N. (2003).
\newblock Relativity in the {G}lobal {P}ositioning {S}ystem.
\newblock {\em Living Reviews in Relativity\/}~{\em 6}, 1.

\bibitem[\protect\citeauthoryear{Banks}{Banks}{2008}]{banksQFT}
Banks, T. (2008).
\newblock {\em Modern Quantum Field Theory: a Concise Introduction}.
\newblock Cambridge: Cambridge University Press.

\bibitem[\protect\citeauthoryear{Belot}{Belot}{2018}]{belotelvis}
Belot, G. (2018).
\newblock Fifty million {E}lvis fans can't be wrong.
\newblock {\em Nous\/}~{\em 52}, 946--981.

\bibitem[\protect\citeauthoryear{Berges}{Berges}{2004}]{bergesqftreview}
Berges, J. (2004).
\newblock Introduction to nonequilibrium quantum field theory.
\newblock {\em AIP Conference Proceedings\/}~{\em 739}, 3.

\bibitem[\protect\citeauthoryear{Bhattacharya, Mohanty, and
  Nautiyal}{Bhattacharya et~al.}{2006}]{bhattacharya-graviton-background}
Bhattacharya, K., S.~Mohanty, and A.~Nautiyal (2006).
\newblock Enhanced polarization of the cosmic microwave background radiation
  from thermal gravitational waves.
\newblock {\em Physical Review Letters\/}~{\em 97}, 251301.

\bibitem[\protect\citeauthoryear{Bianchi and Rovelli}{Bianchi and
  Rovelli}{2010}]{bianchirovelliconstant}
Bianchi, E. and C.~Rovelli (2010).
\newblock Why all these prejudices against a constant?
\newblock https://arxiv.org/abs/1002.3966.

\bibitem[\protect\citeauthoryear{Binney and Tremaine}{Binney and
  Tremaine}{2008}]{binneytremaine}
Binney, J. and S.~Tremaine (2008).
\newblock {\em Galactic Dynamics\/} (2nd ed.).
\newblock Princeton: Princeton University Press.

\bibitem[\protect\citeauthoryear{Bonitz}{Bonitz}{2016}]{bonitz}
Bonitz, M. (2016).
\newblock {\em Quantum Kinetic Theory\/} (2nd ed.).
\newblock Heidelberg: Springer.

\bibitem[\protect\citeauthoryear{Burgess}{Burgess}{2004}]{burgess-eft}
Burgess, C. (2004).
\newblock Quantum gravity in everyday life: General relativity as an effective
  field theory.
\newblock {\em Living Reviews in Relativity\/}~{\em 5}, 7.

\bibitem[\protect\citeauthoryear{Burgess}{Burgess}{2013}]{burgessccp}
Burgess, C. (2013).
\newblock The cosmological constant problem: Why it's hard to get dark energy
  from microphysics.
\newblock https://arxiv.org/abs/1309.4133.

\bibitem[\protect\citeauthoryear{Butterfield and Isham}{Butterfield and
  Isham}{2001}]{butterfieldishamchallenge}
Butterfield, J. and C.~Isham (2001).
\newblock Spacetime and the philosophical challenge of quantum gravity.
\newblock In C.~Callender and N.~Huggett (Eds.), {\em Physics Meets Philosophy
  at the {P}lanck Scale: Contemporary Theories in Quantum Gravity}, pp.\
  33--89. Cambridge, UK: Cambridge University Press.

\bibitem[\protect\citeauthoryear{Butterfield and Isham}{Butterfield and
  Isham}{1999}]{butterfieldishamtime}
Butterfield, J. and C.~J. Isham (1999).
\newblock On the emergence of time in quantum gravity.
\newblock In J.~Butterfield (Ed.), {\em Arguments of Time}, pp.\  111--168.
  Oxford University Press.

\bibitem[\protect\citeauthoryear{Callender and Huggett}{Callender and
  Huggett}{2001}]{callenderhuggettintroduction}
Callender, C. and N.~Huggett (2001).
\newblock Introduction.
\newblock In C.~Callender and N.~Huggett (Eds.), {\em Physics meets philosophy
  at the {P}lanck scale: Contemporary theories in quantum gravity}, pp.\
  1--32. Cambridge, UK: Cambridge University Press.

\bibitem[\protect\citeauthoryear{Calzetta and Hu}{Calzetta and
  Hu}{2008}]{calzettahubook}
Calzetta, E.~A. and B.-L.~B. Hu (2008).
\newblock {\em Nonequilibrium Quantum Field Theory}.
\newblock Cambridge: Cambridge University Press.

\bibitem[\protect\citeauthoryear{Carroll}{Carroll}{2001}]{carrollccp}
Carroll, S. (2001).
\newblock The cosmological constant.
\newblock {\em Living Reviews in Relativity\/}~{\em 4}, 1.

\bibitem[\protect\citeauthoryear{Carroll}{Carroll}{2003}]{carrollgr}
Carroll, S. (2003).
\newblock {\em Spacetime and Geometry: an Introduction to General Relativity}.
\newblock San Francisco, CA.: Addison Wesley.

\bibitem[\protect\citeauthoryear{Cartwright}{Cartwright}{1983}]{Cartwright1983}
Cartwright, N. (1983).
\newblock {\em How the Laws of Physics Lie}.
\newblock Oxford: Oxford University Press.

\bibitem[\protect\citeauthoryear{Christodoulou and Rovelli}{Christodoulou and
  Rovelli}{2018}]{christodoulourovellidetectinggravity}
Christodoulou, M. and C.~Rovelli (2018).
\newblock On the possibility of laboratory evidence for quantum superposition
  of geometries.
\newblock {\em Physics Letters B\/}~{\em 792}, 94--98.

\bibitem[\protect\citeauthoryear{Coleman and Weinberg}{Coleman and
  Weinberg}{1973}]{colemanweinberg}
Coleman, S. and E.~Weinberg (1973).
\newblock Radiative corrections as the origin of spontaneous symmetry breaking.
\newblock {\em Physical Review D\/}~{\em 7}, 1888.

\bibitem[\protect\citeauthoryear{Coleman~Miller and Yunes}{Coleman~Miller and
  Yunes}{2019}]{ligoreview}
Coleman~Miller, M. and N.~Yunes (2019).
\newblock The new frontier of gravitational waves.
\newblock {\em Nature\/}~{\em 568}, 469--476.

\bibitem[\protect\citeauthoryear{Crowther and Linnemann}{Crowther and
  Linnemann}{2017}]{crowtherlinnemann}
Crowther, K. and N.~Linnemann (2017).
\newblock Renormalizability, fundamentality, and a final theory: The role of
  {UV}-completion in the search for quantum gravity.
\newblock {\em British Journal for the Philosophy of Science\/}~{\em 70},
  377--406.

\bibitem[\protect\citeauthoryear{Di{\'o}si}{Di{\'o}si}{1987}]{diosi1987}
Di{\'o}si, L. (1987).
\newblock A universal master equation for the gravitational violations of
  quantum mechanics.
\newblock {\em Physics Letters\/}~{\em 120}, 377--381.

\bibitem[\protect\citeauthoryear{Di{\'o}si}{Di{\'o}si}{2014}]{diosi2014}
Di{\'o}si, L. (2014).
\newblock Gravity related spontaneous wavefunction collapse in bulk matter.
\newblock {\em New Journal of Physics\/}~{\em 16}, 105006.

\bibitem[\protect\citeauthoryear{Donadi, Piscicchia, Curceanu, Di{\'o}si,
  Laubenstein, and Bassi}{Donadi et~al.}{2021}]{donadi2021}
Donadi, S., K.~Piscicchia, C.~Curceanu, L.~Di{\'o}si, M.~Laubenstein, and
  A.~Bassi (2021).
\newblock Underground test of gravity-related wavefunction collapse.
\newblock {\em Nature Physics\/}~{\em 17}, 74--78.

\bibitem[\protect\citeauthoryear{Donoghue}{Donoghue}{1997}]{donoghue-eft-1997}
Donoghue, J.~F. (1997).
\newblock Introduction to the effective field theory description of gravity.
\newblock In F.~Cornet and M.~J. Herrero (Eds.), {\em Proceedings of the
  Advanced School on Effective Field Theories, Almunecar, Spain}. World
  Scientific.
\newblock arxiv.org/gr-qc/9512024.

\bibitem[\protect\citeauthoryear{Donoghue}{Donoghue}{2012}]{donoghue-eft-2012}
Donoghue, J.~F. (2012).
\newblock The effective field theory treatment of quantum gravity.
\newblock {\em AIP Conference Proceedings\/}~{\em 1483}, 73.

\bibitem[\protect\citeauthoryear{Donoghue, Golowich, and Holstein}{Donoghue
  et~al.}{2014}]{dynamicsofstandardmodel}
Donoghue, J.~F., E.~Golowich, and B.~R. Holstein (2014).
\newblock {\em Dynamics of the Standard Model (Second Edition)}.
\newblock Cambridge: Cambridge University Press.

\bibitem[\protect\citeauthoryear{Fuchs and Peres}{Fuchs and
  Peres}{2000}]{fuchsperes}
Fuchs, C. and A.~Peres (2000).
\newblock Quantum theory needs no ``interpretation''.
\newblock {\em Physics Today\/}~{\em 53\/}(3), 70--71.

\bibitem[\protect\citeauthoryear{Fuchs, Mermin, and Schack}{Fuchs
  et~al.}{2014}]{fuchsmerminschack}
Fuchs, C.~A., N.~D. Mermin, and R.~Schack (2014).
\newblock An introduction to {QB}ism with an application to the locality of
  quantum mechanics.
\newblock {\em American Journal of Physics\/}~{\em 82}, 749--754.

\bibitem[\protect\citeauthoryear{Gell-Mann and Hartle}{Gell-Mann and
  Hartle}{1989}]{gellmannhartle}
Gell-Mann, M. and J.~B. Hartle (1989).
\newblock {Q}uantum {M}echanics in the {L}ight of {Q}uantum {C}osmology.
\newblock In W.~H. Zurek (Ed.), {\em Complexity, Entropy and the Physics of
  Information}, pp.\  425--459. Redwood City, California: Westview Press.

\bibitem[\protect\citeauthoryear{Gell-Mann and Hartle}{Gell-Mann and
  Hartle}{1993}]{gellmannhartle93}
Gell-Mann, M. and J.~B. Hartle (1993).
\newblock Classical equations for quantum systems.
\newblock {\em Physical Review D\/}~{\em 47}, 3345--3382.

\bibitem[\protect\citeauthoryear{Griffiths}{Griffiths}{1993}]{Griffiths1993}
Griffiths, R. (1993).
\newblock Consistent interpretation of quantum mechanics using quantum
  trajectories.
\newblock {\em Physical Review Letters\/}~{\em 70}, 2201--2204.

\bibitem[\protect\citeauthoryear{Griffiths}{Griffiths}{1984}]{griffiths}
Griffiths, R.~B. (1984).
\newblock Consistent histories and the interpretation of quantum mechanics.
\newblock {\em Journal of Statistical Physics\/}~{\em 36}, 219--272.

\bibitem[\protect\citeauthoryear{Halliwell}{Halliwell}{1995}]{halliwellreview}
Halliwell, J.~J. (1995).
\newblock A review of the decoherent histories approach to quantum mechanics.
\newblock {\em Annals of the New York Academy of Sciences\/}~{\em 755},
  726--740.

\bibitem[\protect\citeauthoryear{Hawking}{Hawking}{1979}]{hawking-centenary}
Hawking, S. (1979).
\newblock The path-integral approach to quantum gravity.
\newblock In S.~Hawking and W.~Israel (Eds.), {\em General Relativity: an
  {E}instein centenary survey}, pp.\  746--789. Cambridge: Cambridge University
  Press.

\bibitem[\protect\citeauthoryear{Healey}{Healey}{2012}]{healeypragmatism}
Healey, R. (2012).
\newblock Quantum theory: A pragmatist approach.
\newblock {\em British Journal for the Philosophy of Science\/}~{\em 63},
  729--771.

\bibitem[\protect\citeauthoryear{Healey}{Healey}{2017}]{healeypragmatismbook}
Healey, R. (2017).
\newblock {\em The Quantum Revolution in Philosophy}.
\newblock Oxford: Oxford University Press.

\bibitem[\protect\citeauthoryear{Hossenfelder}{Hossenfelder}{2018}]{hossenfeldernaturalness}
Hossenfelder, S. (2018).
\newblock Screams for explanation: Finetuning and naturalness in the
  foundations of physics.
\newblock https://arxiv.org/abs/1801.02176.

\bibitem[\protect\citeauthoryear{Hu, Roura, and Verdaguer}{Hu
  et~al.}{2004}]{hu-roura-verdaguer}
Hu, B., A.~Roura, and E.~Verdaguer (2004).
\newblock Induced quantum metric fluctuations and the validity of semiclassical
  gravity.
\newblock {\em Physical Review D\/}~{\em 70}, 044002.

\bibitem[\protect\citeauthoryear{Hu and Verdaguer}{Hu and
  Verdaguer}{2008}]{hu-verdaguer-review}
Hu, B. and E.~Verdaguer (2008).
\newblock Stochastic gravity: Theory and applications.
\newblock {\em Living Reviews in Relativity\/}~{\em 11}, 3.

\bibitem[\protect\citeauthoryear{Keller, Wellenhofer, Hebeler, and
  Schwenk}{Keller et~al.}{2021}]{kellerneutronmatter}
Keller, J., C.~Wellenhofer, K.~Hebeler, and A.~Schwenk (2021).
\newblock Neutron matter at finite temperature based on chiral effective field
  theory interactions.
\newblock {\em Physical Review C\/}~{\em 103}, 055806.

\bibitem[\protect\citeauthoryear{Kiefer}{Kiefer}{2007}]{kieferqg}
Kiefer, C. (2007).
\newblock {\em Quantum Gravity\/} (2nd ed.).
\newblock Oxford, UK: Oxford University Press.

\bibitem[\protect\citeauthoryear{Kippenhahn, Weigert, and Weiss}{Kippenhahn
  et~al.}{2012}]{kippenhahn}
Kippenhahn, R., A.~Weigert, and A.~Weiss (2012).
\newblock {\em Stellar Structure and Evolution\/} (2nd ed.).
\newblock New York/Dordrecht/London: Springer.

\bibitem[\protect\citeauthoryear{Koberinski}{Koberinski}{2017}]{koberinskicosmologicalconstant}
Koberinski, A. (2017).
\newblock Problems with the cosmological constant problem.
\newblock http://philsci-archive.pitt.edu/14244/.

\bibitem[\protect\citeauthoryear{Koberinski}{Koberinski}{2021}]{koberinskivacuum}
Koberinski, A. (2021).
\newblock Regularizing (away) vacuum energy.
\newblock {\em Foundations of Physics\/}~{\em 51}, 20.

\bibitem[\protect\citeauthoryear{Kuhn}{Kuhn}{1962}]{kuhn}
Kuhn, T. (1962).
\newblock {\em The Structure of Scientific Revolutions}.
\newblock Chicago: University of Chicago Press.

\bibitem[\protect\citeauthoryear{Li}{Li}{2013}]{li-qed}
Li, B. (2013).
\newblock Interpretive stratagies for deductively insecure theories: The case
  of early quantum electrodynamics.
\newblock {\em Studies in History and Philosophy of Modern Physics\/}~{\em 44},
  395--403.

\bibitem[\protect\citeauthoryear{Marletto and Vedral}{Marletto and
  Vedral}{2017}]{marlettovedralgravity}
Marletto, C. and V.~Vedral (2017).
\newblock Witness gravity's quantum side in the lab.
\newblock {\em Nature\/}~{\em 547}, 156--158.

\bibitem[\protect\citeauthoryear{Mattingly}{Mattingly}{2005}]{mattinglynecessary}
Mattingly, J. (2005).
\newblock Is quantum gravity necessary.
\newblock In A.~Kox and J.~Eisenstaedt (Eds.), {\em The Universe of General
  Relativity}, pp.\  327--328. Boston: Birkh{\"a}user.

\bibitem[\protect\citeauthoryear{Mattingly}{Mattingly}{2009}]{mattinglymongrel}
Mattingly, J. (2009).
\newblock Mongrel gravity.
\newblock {\em Erkenntnis\/}~{\em 70}, 379--395.

\bibitem[\protect\citeauthoryear{Mattingly}{Mattingly}{2014}]{mattinglyunprincipled}
Mattingly, J. (2014).
\newblock Unprincipled microgravity.
\newblock {\em Studies in History and Philosophy of Modern Physics\/}~{\em 46},
  179--185.

\bibitem[\protect\citeauthoryear{Niedermaier}{Niedermaier}{2007}]{niedermaier}
Niedermaier, M. (2007).
\newblock The asymptotic safety scenario in quantum gravity - an introduction.
\newblock {\em Classical and Quantum Gravity\/}~{\em 24}, R171--230.

\bibitem[\protect\citeauthoryear{Omnes}{Omnes}{1988}]{omnes}
Omnes, R. (1988).
\newblock Logical reformulation of quantum mechanics. i. foundations.
\newblock {\em Journal of Statistical Physics\/}~{\em 53}, 893--932.

\bibitem[\protect\citeauthoryear{Omnes}{Omnes}{1992}]{omnes92}
Omnes, R. (1992).
\newblock Consistent interpretations of quantum mechanics.
\newblock {\em Reviews of Modern Physics\/}~{\em 64}, 339--382.

\bibitem[\protect\citeauthoryear{Padilla}{Padilla}{2015}]{padillaccp}
Padilla, A. (2015).
\newblock Lectures on the cosmological constant problem.
\newblock https://arxiv.org/abs/1502.05296.

\bibitem[\protect\citeauthoryear{Padmanabhan}{Padmanabhan}{2014}]{padmanabhan-boundary}
Padmanabhan, T. (2014).
\newblock A short note on the boundary term for the {H}ilbert action.
\newblock {\em Modern Physics Letters A\/}~{\em 29}, 1450037.

\bibitem[\protect\citeauthoryear{Page and Geilker}{Page and
  Geilker}{1981}]{Page1981}
Page, D.~N. and C.~D. Geilker (1981).
\newblock Indirect evidence for quantum gravity.
\newblock {\em Physical Review Letters\/}~{\em 47}, 979--982.

\bibitem[\protect\citeauthoryear{Penrose}{Penrose}{1969}]{penrose1969}
Penrose, R. (1969).
\newblock Gravitational collapse: The role of general relativity.
\newblock {\em Il Nuovo Cimento\/}~{\em Numero Speziale I}, 257.
\newblock Reprinted in \emph{General Relativity and Gravitation} 34 (2002)
  pp.1141--1165.

\bibitem[\protect\citeauthoryear{Penrose}{Penrose}{1996}]{penrose-1996}
Penrose, R. (1996).
\newblock On gravity's role in quanutm state reduction.
\newblock {\em General Relativity and Gravitation\/}~{\em 28}, 581--600.

\bibitem[\protect\citeauthoryear{Penrose}{Penrose}{2014}]{penrose-2014}
Penrose, R. (2014).
\newblock On the gravitization of quantum mechanics 1: Quantum state reduction.
\newblock {\em Foundations of Physics\/}~{\em 44}, 557--575.

\bibitem[\protect\citeauthoryear{Peskin and Schroeder}{Peskin and
  Schroeder}{1995}]{peskinschroeder}
Peskin, M.~E. and D.~V. Schroeder (1995).
\newblock {\em An introduction to Quantum Field Theory}.
\newblock Reading, Massachusetts: Addison-Wesley.

\bibitem[\protect\citeauthoryear{Pinto-Neto and Struyve}{Pinto-Neto and
  Struyve}{2018}]{pinto-neto-struyve-gravity}
Pinto-Neto, N. and W.~Struyve (2018).
\newblock {B}ohmian quantum gravity and cosmology.
\newblock https://arxiv.org/abs/1801.03353.

\bibitem[\protect\citeauthoryear{Polchinski}{Polchinski}{2006}]{polchinskiccp}
Polchinski, J. (2006).
\newblock The cosmological constant and the string landscape.
\newblock https://arxiv.org/abs/hep-th/0603249.

\bibitem[\protect\citeauthoryear{Reuter and Saueressig}{Reuter and
  Saueressig}{2012}]{reutersaueressig}
Reuter, M. and F.~Saueressig (2012).
\newblock Quantum {E}instein gravity.
\newblock https://arxiv.org/abs/1202.2274.

\bibitem[\protect\citeauthoryear{Rickles}{Rickles}{2008}]{ricklesQGreview}
Rickles, D. (2008).
\newblock Quantum gravity: A primer for philosophers.
\newblock In D.~Rickles (Ed.), {\em The Ashgate Companion to Contemporary
  Philosophy of Physics}, pp.\  262--365. New York: Ashgate.

\bibitem[\protect\citeauthoryear{Rovelli}{Rovelli}{1996}]{rovelli-relational}
Rovelli, C. (1996).
\newblock Relational quantum mechanics.
\newblock {\em International Journal of Theoretical Physics\/}~{\em 35}, 1637.
\newblock Substantially revised version at arxiv.org/abs/quant-ph/9609002.

\bibitem[\protect\citeauthoryear{Rovelli}{Rovelli}{2008}]{rovelli-scholarpedia}
Rovelli, C. (2008).
\newblock Quantum gravity.
\newblock {\em Scholarpedia\/}~{\em 3}, 7117.
\newblock www.scholarpedia.org/article/Quantum\_gravity.

\bibitem[\protect\citeauthoryear{Rugh and Zinkernagel}{Rugh and
  Zinkernagel}{2002}]{rughzinkernagel}
Rugh, S.~E. and H.~Zinkernagel (2002).
\newblock The quantum vacuum and the cosmological constant problem.
\newblock {\em Studies in History and Philosophy of Modern Physics\/}~{\em 33},
  663--705.

\bibitem[\protect\citeauthoryear{Saunders}{Saunders}{2002}]{saunderscosmologicalconstant}
Saunders, S. (2002).
\newblock Is the zero-point energy real?
\newblock In M.~Kuhlmann, H.~Lyre, and A.~Wayne (Eds.), {\em Ontological
  Aspects of Quantum Field Theory}, pp.\  313--344. Singapore: World
  Scientific.

\bibitem[\protect\citeauthoryear{Schneider}{Schneider}{2020}]{schneiderccp}
Schneider, M.~D. (2020).
\newblock What's the problem with the cosmological constant?
\newblock {\em Philosophy of Science\/}~{\em 87}, 1--20.

\bibitem[\protect\citeauthoryear{Smeenk}{Smeenk}{2017}]{smeenk-testinginflation}
Smeenk, C. (2017).
\newblock Testing inflation.
\newblock In K.~Chamchan, J.~Barrow, and S.~Saunders (Eds.), {\em Philosophy of
  Cosmology}. Cambridge: Cambridge University Press.

\bibitem[\protect\citeauthoryear{Stamp}{Stamp}{2015}]{stampqg}
Stamp, P. (2015).
\newblock Rationale for a correlated worldline theory of quantum gravity.
\newblock {\em New Journal of Physics\/}~{\em 17}, 065017.

\bibitem[\protect\citeauthoryear{Susskind and Lindesay}{Susskind and
  Lindesay}{2005}]{susskindlindesay}
Susskind, L. and J.~Lindesay (2005).
\newblock {\em An introduction to black holes, information and the string
  theory revolution: the Holographic Universe}.
\newblock Singapore: World Scientific.

\bibitem[\protect\citeauthoryear{Urena-Lopez}{Urena-Lopez}{2019}]{scalarfielddarkmatter-review}
Urena-Lopez, L.~A. (2019).
\newblock Brief review on scalar field dark matter models.
\newblock {\em Frontiers of Astronomy and Space Science\/}.
\newblock https://doi.org/10.3389/fspas.2019.00047.

\bibitem[\protect\citeauthoryear{Wald}{Wald}{1984}]{waldrelativitybook}
Wald, R.~M. (1984).
\newblock {\em General Relativity}.
\newblock Chicago: University of Chicago Press.

\bibitem[\protect\citeauthoryear{Wall}{Wall}{2012}]{wallgsl}
Wall, A.~C. (2012).
\newblock A proof of the generalized second law for rapidly changing fields and
  arbitrary horizon slices.
\newblock {\em Physical Review D\/}~{\em 85}, 104049.
\newblock Most recent version at https://arxiv.org/abs/1105.3445v5.

\bibitem[\protect\citeauthoryear{Wallace}{Wallace}{2011}]{wallacecritique}
Wallace, D. (2011).
\newblock Taking particle physics seriously: a critique of the algebraic
  approach to quantum field theory.
\newblock {\em Studies in the History and Philosophy of Modern Physics\/}~{\em
  42}, 116--125.

\bibitem[\protect\citeauthoryear{Wallace}{Wallace}{2012}]{wallacebook}
Wallace, D. (2012).
\newblock {\em The Emergent Multiverse: Quantum Theory according to the
  {E}verett Interpretation}.
\newblock Oxford: Oxford University Press.

\bibitem[\protect\citeauthoryear{Wallace}{Wallace}{2016a}]{wallacecosmology}
Wallace, D. (2016a).
\newblock Interpreting the quantum mechanics of cosmology.
\newblock Forthcoming in ``Guide to the Philosophy of Cosmology'', Ijjas and
  Loewer (ed.) Preprint at http://philsci-archive.pitt.edu/15295/.

\bibitem[\protect\citeauthoryear{Wallace}{Wallace}{2016b}]{wallacequantumstatmech}
Wallace, D. (2016b).
\newblock Probability and irreversibility in modern statistical mechanics:
  Classical and quantum.
\newblock To appear in D. Bedingham, O. Maroney and C. Timpson (eds.),
  \emph{Quantum Foundations of Statistical Mechanics} (Oxford University Press,
  forthcoming). Preprint at https://arxiv.org/abs/2104.11223.

\bibitem[\protect\citeauthoryear{Wallace}{Wallace}{2019a}]{wallace-isolated-2}
Wallace, D. (2019a).
\newblock Isolated systems and their symmetries, part {II}: Local and global
  symmetries of field theories.
\newblock Forthcoming in \emph{Studies in the History and Philosophy of
  Science}. Preprint at http://philsci-archive.pitt.edu/19729/.

\bibitem[\protect\citeauthoryear{Wallace}{Wallace}{2019b}]{wallacenaturalness}
Wallace, D. (2019b).
\newblock Naturalness and emergence.
\newblock {\em Monist\/}~{\em 102}, 499--524.

\bibitem[\protect\citeauthoryear{Wallace}{Wallace}{2019c}]{wallaceorthodoxy}
Wallace, D. (2019c).
\newblock What is orthodox quantum mechanics?
\newblock In A.~Cordero (Ed.), {\em Philosophers Look at Quantum Mechanics},
  pp.\  285--309. Switzerland: Springer.

\bibitem[\protect\citeauthoryear{Weinberg}{Weinberg}{1979}]{weinbergasymptotic}
Weinberg, S. (1979).
\newblock Ultraviolet divergences in quantum theories of gravitation.
\newblock In S.~Hawking and W.~Israel (Eds.), {\em General Relativity: an
  {E}instein Centenary Survey}, pp.\  790--832. Cambridge: Cambridge University
  Press.

\bibitem[\protect\citeauthoryear{Weinberg}{Weinberg}{1989}]{weinbergcosmologicalconstant}
Weinberg, S. (1989).
\newblock The cosmological constant problem.
\newblock {\em Reviews of Modern Physics\/}~{\em 61}, 1--23.

\bibitem[\protect\citeauthoryear{Weinberg}{Weinberg}{1995}]{weinbergqft2}
Weinberg, S. (1995).
\newblock {\em The Quantum Theory of Fields, Volume II: Modern Applications}.
\newblock Cambridge: Cambridge University Press.

\bibitem[\protect\citeauthoryear{Weinberg}{Weinberg}{2008}]{weinbergcosmology}
Weinberg, S. (2008).
\newblock {\em Cosmology}.
\newblock Oxford: Oxford University Press.

\bibitem[\protect\citeauthoryear{Wiltshire}{Wiltshire}{2011}]{wiltshirewhatisdust}
Wiltshire, D.~L. (2011).
\newblock What is dust? - physics foundations of the averaging problem in
  cosmology.
\newblock {\em Classical and Quantum Gravity\/}~{\em 28}, 164006.

\bibitem[\protect\citeauthoryear{Zurek and Paz}{Zurek and
  Paz}{1999}]{zurekpazplanetary}
Zurek, W.~H. and J.~P. Paz (1999).
\newblock Why we donâ€™t need quantum planetary dynamics: Decoherence and the
  correspondence principle for chaotic systems.
\newblock In D.~Greenberger, W.~L. Reiter, and A.~Zeilinger (Eds.), {\em
  Epistemological and Experimental Perspectives on Quantum Physics}. Dordrecht:
  Springer.
\newblock https://arxiv.org/abs/quant-ph/9612037.

\end{thebibliography}

\end{document}